# Continuous quantification of viral plaque dynamics using ultra-large-area label-free imaging enables rapid antiviral susceptibility testing

Merve Eryilmaz[†,1,2], Yuzhu Li[†,1,2,3], Xiao Wang[1], Max Zhang[4], Alp Inegol[1], Zixiang Ji[5], Lucas Thai[5], Guangdong Ma[1], Akihiko Fujisawa[6], Kazunori Yamaguchi[7], Aydogan Ozcan*[1,2,3,8]

[1]Department of Electrical and Computer Engineering, University of California, Los Angeles, CA, USA, 90095 [2]Department of Bioengineering, University of California, Los Angeles, CA, USA, 90095 [3]California NanoSystems Institute (CNSI), University of California, Los Angeles, CA, USA, 90095 [4]Department of Biochemistry, University of California, Los Angeles, CA, USA, 90095 [5]Department of Computer Science, University of California, Los Angeles, CA, USA, 90095 [6]JDI Display America, Inc., San Jose, CA 95110, USA [7]Sensor Department, Application Engineering Division, Japan Display Inc., Tokyo 105-0003, Japan [8]Department of Surgery, University of California, Los Angeles, CA, USA, 90095

*Correspondence: Aydogan Ozcan, ozcan@ucla.edu

†Equal contributing authors

**ABSTRACT**

The plaque reduction assay (PRA) remains the gold standard for antiviral susceptibility testing, evaluating drug potency by measuring reductions in plaque-forming units (PFUs). However, the traditional PRA is time-consuming, labor-intensive, prone to manual counting errors, and offers limited scalability. Moreover, its reliance on destructive fixation and chemical staining reduces the assay to a static, endpoint observation, obscuring the dynamic, time-resolved kinetics of dose-dependent viral inhibition. Here, we introduce a label-free, time-resolved PRA platform that transforms the conventional assay into a continuous, high-dimensional measurement of viral infection dynamics. Our system integrates a compact lens-free imaging setup with a custom-designed ultra-large-area (100 cm²) thin-film transistor (TFT) image sensor and deep learning-based algorithms to autonomously quantify PFU dynamics within an incubator. Validated using herpes simplex virus type-1 (HSV-1) treated with acyclovir, the platform matched chemically-stained ground truth measurements with zero false positives while accelerating readout by ~26 hours. Crucially, our system revealed that increasing drug concentrations induce temporally distinct delays and suppress new PFU formation, enabling conclusive drug efficacy evaluations within ~60 hours post-infection. This scalable, label-free framework redefines antiviral susceptibility testing as a rapid, time-resolved and information-rich measurement framework, providing a generalizable platform for virology research, high-throughput drug screening, and clinical diagnostics.

## INTRODUCTION

Antiviral drugs are pharmacological agents designed to selectively inhibit specific stages of the viral replication cycle, including viral attachment, entry, genome replication, assembly, and release, while minimizing toxicity to host cells[1,2]. Due to their narrow target specificity and the high attrition rate among candidate compounds, antiviral drug development follows a rigorous, multistage pipeline[3-5]. This process begins with *in vitro* antiviral screening, in which compounds are initially assessed for their capacity to suppress viral replication in cell culture systems[6,7]. Promising candidates then advance to *in vivo* efficacy evaluations in animal models, followed by comprehensive toxicology assessments, and ultimately to clinical trials[3,7,8]; see **Fig. 1a**.

Within the established antiviral drug development framework, the plaque reduction assay (PRA) remains the widely-accepted gold standard for assessing the antiviral activity of compounds against a broad range of viruses.[3,7-11] As a frequently-used *in vitro* antiviral screening method[7], the PRA quantifies antiviral potency by directly measuring the reduction in plaque-forming units (PFUs) in drug-treated samples relative to untreated controls[3,7,12], thereby helping prioritize promising candidates for further preclinical evaluation. As illustrated in **Fig. 1b**, the assay involves seeding a confluent cell monolayer, inoculating it with a standardized viral inoculum, and overlaying the infected monolayer with a semi-solid medium containing serial dilutions of the candidate antiviral agent. After an incubation period sufficient for plaques to grow to visible size under manual inspection, the monolayer is fixed and stained with crystal violet to reveal a clear contrast between the PFUs and the uninfected cells, allowing the number of PFUs to be manually counted[9,12]. Accordingly, the PRA fulfills three major functions in antiviral drug discovery and development: (i) *in vitro* screening and verification of antiviral activity[13]; (ii) dose–response analysis to derive potency metrics[7,10]; and (iii) phenotypic monitoring of antiviral resistance through comparative susceptibility testing of clinical or patient-derived viral isolates[9,14].

However, despite being the gold standard method, the conventional PRA remains constrained by its dependence on endpoint plaque visualization, requiring laborious post-incubation fixation, staining, and manual counting by an expert. To ensure that plaques grow to a macroscopically visible size before staining, manual visualization and counting, the assay generally requires multi-day incubation, rendering the traditional workflow time-consuming[9,11,12]. Moreover, the use of chemical dyes and toxic fixatives further raises health and environmental concerns. As a result of this chemical staining process, the conventional PRA is largely restricted to an endpoint readout rather than a continuous measurement, fundamentally limiting its capacity to resolve the dynamic, longitudinal progression of infection[3,7,9,10,15]. Furthermore, reliance on a single static endpoint can compromise quantitative accuracy: adjacent PFUs may expand and merge into dense clusters, making the differentiation of individual plaques within the clusters unreliable based on static endpoint morphological information. Together, these limitations necessitate a paradigm shift in virological assays—moving beyond static, labor-intensive manual counting toward autonomous, high-dimensional phenotyping that captures the continuous spatio-temporal evolution of infection dynamics in real-time.

To bypass some of these drawbacks, alternative approaches such as molecular assays[16-21] (e.g., qPCR-based viral load quantification) and rapid dye-uptake assays[21,22] (DUA) are frequently employed. However, a fundamental limitation of molecular techniques is their inability to distinguish between infectious virions and defective or inactivated viral particles. Furthermore, these methods typically necessitate prior viral isolation, propagation, and sometimes labeling, which can artifactually overestimate antiviral efficacy due to altered viral fitness or incomplete representation of the original isolate. Fluorescence-based platforms[23-25], including immunofluorescence focus assays (FFAs) and reporter virus systems[26,27] (e.g., those expressing fluorescent proteins), generally depend on exogenous dyes, antibodies, or molecular substrates for detection.[28-31] These approaches often require specialized optical instrumentation, substantially increasing reagent and equipment costs and thereby constraining their scalability and widespread adoption. More recent label-free technologies, such as holographic scanners, live-cell analysis platforms, impedance-based measurements (e.g., real-time cell analysis systems), and flow cytometry, have expanded the antiviral-screening toolkit by enabling faster, more sensitive, and often real-time assessment of infection-associated cellular responses[11,32-40]. Nevertheless, despite these advances, existing platforms have not yet minimized assay setup and analysis complexity or achieved cost-effectiveness for routine high-throughput applications. Although recent studies have successfully employed machine learning to automate the detection of PFUs and virus-induced cytopathic effects[41-44], these approaches have primarily been developed for infection detection, plaque quantification, or phenotypic classification rather than for fully automated susceptibility testing with time-resolved dose–response profiling, leaving an important gap in dynamic drug-efficacy assessment.

Here, we introduce a time-resolved measurement framework that captures the full temporal evolution of plaque formation and viral propagation, enabling continuous quantification of viral replication kinetics under drug pressure. This transforms antiviral susceptibility testing from an endpoint assay into a dynamic phenotyping framework. This high-throughput, label-free and dynamic PRA platform is automated to run within an incubator, saving time, labor and cost. By preserving the core PRA workflow while replacing laborious and subjective manual steps with deep-learning-driven spatio-temporal analysis, this autonomous system enables the precise, dynamic monitoring of dose-dependent viral inhibition (**Fig. 1**). The platform is built upon a customized, compact lens-free imaging setup (see **Fig. 1b** and **Supplementary Video 1**) designed to fit inside a standard incubator. This system utilizes a custom-designed thin-film-transistor (TFT) image sensor with an ultra-large active area of 100 $cm^2$, enabling snapshot capture of a large sample area (~15.4 $cm^2$ per lens-free image). Once a sample is prepared, it is placed onto our device for continuous time-lapse imaging at 1-hour intervals during the incubation period. These time-lapse raw lens-free images are then transformed into differential image stacks, which amplify the dynamic expansion of PFUs and provide optimized inputs for our custom-designed PFU detection neural network. By learning the unique spatio-temporal features of viral plaque growth, the network generates whole-well probability maps (see **Supplementary Video 2**) and binary detection masks, facilitating automated, longitudinal extraction of four key quantitative metrics that represent important dimensions of infection dynamics, including total PFU count $N$, total PFU area coverage $A$, new PFU formation $\Delta N$, and incremental PFU area growth $\Delta A$. The efficacy of this system was rigorously validated against chemically stained ground truths obtained after 96 hours of incubation using the traditional PRA workflow. When tested on HSV-1 (herpes simplex virus type-1) infected, untreated wells (positive controls without an antiviral drug) as well as uninfected, untreated wells (negative controls), our system matched the traditional endpoint 96-hour results with zero false positives across all time points (see e.g., **Supplementary Video 3**). Crucially, the system detected more than 90% of PFUs in positive controls within 70 hours—accelerating the readout by ~26 hours compared to the standard incubation period. Moreover, our platform resolved dense PFU clusters prior to their coalescence, overcoming a major quantification bottleneck inherent to static, endpoint-based PFU staining assays.

We evaluated the system's capacity for drug efficacy assessment using HSV-1 infected, acyclovir (ACV) treated wells spanning different concentrations of this FDA-approved antiviral compound. The platform successfully quantified concentration-dependent viral inhibition: increasing ACV dosage (e.g., 1.02, 2.04, and 3.06 µg/mL) progressively delayed and suppressed the peak values of $\Delta N$ and $\Delta A$, culminating in near-complete viral suppression at 5.12 µg/mL. Even at maximum inhibition with high ACV dosing, the system still detected small-scale PFUs (<200 µm in diameter) at 90 hours, which remain difficult to identify to the human eye even after chemical staining following the 96-hour incubation period. Importantly, by analyzing the temporal delays and suppression of $\Delta N$ and $\Delta A$ peak values, conclusive drug efficacy was determined in ~60 hours post-infection. Therefore, our approach enables early prediction of antiviral efficacy by observing the trajectory of infection dynamics, allowing conclusive assessment based on the temporal evolution of plaque formation rather than its final state. Because the label-free method relies on generic spatio-temporal signatures of cytopathic progression rather than virus-specific labels or markers, it is inherently adaptable across viral systems. By providing accurate, dynamic monitoring of antiviral susceptibility testing in a fully automated, label-free manner, our computational PRA platform has the potential to be widely applied to accelerate antiviral drug discovery and resistance profiling across various viral pathogens.

## RESULTS

### Label-free and automated PRA system design

To demonstrate the efficacy of our label-free, rapid, and automated PRA system, our experimental design encompassed three distinct conditions: (1) HSV-1 infected, ACV-untreated wells; (2) HSV-1 infected wells treated with varying ACV concentrations (1.02, 2.04, 3.06, and 5.12 µg/mL), and (3) uninfected, untreated negative control wells. All sample preparation and assay steps followed the standard PRA protocol, as detailed in the *Materials and Methods* section. Once the protocol was completed (culturing Vero Cells, infecting the cells with HSV-1, and agar overlaying), the sample plate was placed into our cost-effective lens-free imaging setup housed within a standard incubator (see **Fig. 2a, b** and **Supplementary Video 1**). This system utilizes a laser diode-based illumination at $\lambda = 515$ nm and a custom-designed TFT image sensor featuring an ultra-large active area of 100 $cm^2$ (pixel size: 100 µm). Benefiting from this extensive active area, image acquisition is performed in a single shot across the entire well, eliminating the need for mechanical scanning of the sample or the sensor. To compensate for the relatively large pixel size of the TFT sensor, the setup employs a design with a lens-free fringe magnification factor[45-47] of 2.55×. This ensures that early-stage PFU features are sufficiently enlarged through diffraction to occupy multiple sensor pixels, thereby enabling sensitive and robust neural network-based identification. Using this label-free lens-free setup, time-lapse imaging was performed by monitoring a single well over 96 hours at 1-hour intervals to capture the spatio-temporal evolution of the samples inside the incubator. At the end of the 96-h experiment, the samples were fixed and chemically stained with crystal violet to establish the ground-truth images, which were used to guide training dataset annotation and validate the system's performance during the blind testing phase.

Computational analysis of the time-lapse data was initiated at the 22$^{nd}$ hour of incubation to allow sufficient temporal data accumulation required for our differential analysis and neural network-based detection pipeline (see the *Materials and Methods* section for details). As illustrated in **Fig. 2c**, after the initial image pre-processing, the aligned raw lens-free images were converted into dual-channel, label-free differential image stacks by subtracting the intensity of the preceding frame from the current frame at two distinct temporal intervals ($\Delta T = 8$ h and 12 h). This differential analysis effectively suppressed the static cellular background (eliminating false positives) while amplifying the dynamic morphological signatures of expanding PFUs within the well area (improving sensitivity and suppressing false negatives). Finally, these dual-channel time-lapse differential images were spatially cropped and temporally segmented into 8-frame sequences, forming 4D spatio-temporal tensors that were fed into the PFU detection neural network for automated PFU detection. It is worth noting that the selection of the dual-channel differential images as the network input was rigorously optimized. Ablation studies demonstrated that omitting either of the two differential channels ($\Delta T$ = 8 h or 12 h) degraded network performance, whereas incorporating additional input channels (e.g., raw intensity images, reconstructed amplitude, and phase channels) did not achieve statistically significant performance improvements and therefore were omitted. Detailed comparative analyses validating this optimal input configuration are provided in **Supplementary Figure 1** and the *Discussion* section.

To train and validate the PFU detection neural network, we prepared 24 HSV-1 infected ACV-untreated wells (19 for training; 5 for validation) and 10 uninfected, untreated negative control wells (5 for training; 5 for validation). Using data augmentation, these incubation experiments yielded 13,267 positive and 44,029 negative 4D spatio-temporal tensors for training, along with 5,051 positive and 9,617 negative tensors reserved for validation. In the training phase, for positive sample curation, we manually identified the centroids of terminal PFUs on differential images at T = 96 h, using the corresponding chemically stained ground-truth images as reference. Leveraging the precise spatial registration obtained during pre-processing, these endpoint coordinates of the PFUs were retrospectively mapped across prior time frames to localize early-stage, low-contrast viral plaques. To generate the negative training dataset with non-PFU "artifacts", centroids were manually selected from PFU-free regions within infected wells, supplemented by random spatial sampling across uninfected control wells. Centered on these spatially verified coordinates, the dual-channel time-lapse differential images were extracted and formatted into 4D spatio-temporal tensors, denoted as $X \in \mathbb{R}^{C \times D \times H \times W}$, to serve as direct input to the PFU neural network. These differential images were cropped to 50×50 pixels (H = 50, W = 50) around the annotated centroids. Temporally, the input spans 8 consecutive time points (D = 8), with each time point comprising the two differential feature channels ($C = 2$, corresponding to $\Delta T$ = 8 h and 12 h). To further optimize the training dataset and reject potential false positives, an iterative hard-negative mining strategy was employed (detailed in the *Materials and Methods* section). By iteratively scanning uninfected negative control wells and integrating false-positive predictions from prior iterations across four cycles, the final training dataset was enriched with challenging examples, particularly those misleading the PFU detection neural network model.

Following training, the PFU detection neural network, custom-built based on a Dense-Net[48] architecture with pseudo-3D[49] convolutional layers, was applied to whole-well blind testing on independent, previously ***unseen*** experimental data. This blind test set consisted of 12 HSV-1 infected, ACV-treated wells containing a total of 300 distinct PFUs, 3 HSV-1 infected, ACV-untreated wells with 113 distinct PFUs (positive controls), and 3 uninfected, untreated wells (negative controls) with ~$2.65\times10^6$ distinct 4D test tensors when accounting for all spatial regions and time points tested within the negative control samples. Notably, since the training dataset was derived exclusively from ACV-untreated samples (both infected and uninfected), evaluating the model on the 12 ACV-treated wells with 300 distinct PFUs constitutes an external test of its generalization ability, as the training and validation data were never exposed to antiviral compounds. During blind inference, the network received local 4D spatio-temporal tensors as input and produced a probability score between 0 and 1, reflecting the likelihood of a PFU being present within a given region. To obtain continuous probability predictions across the entire well, an overlapping sliding-window raster-scanning approach was used, as shown in **Supplementary Video 2**. A series of post-processing steps, including morphological operations, temporal maximum projection, and geometric distortion correction (described in the *Materials and Methods* section), were then applied to refine the resulting probability maps. A decision threshold of 0.5 was applied to convert the final probability maps into binary PFU detection masks, from which the total PFU count $N$, total PFU area coverage $A$, new PFU formation $\Delta N$ and, incremental PFU area growth $\Delta A$ were extracted, providing continuous/dynamic quantitative measurements of viral progression over the incubation period.

**Evaluation of our system on HSV-1 infected, ACV-untreated samples**

**Figure 3** illustrates the performance of our system on a representative HSV-1 infected, ACV-untreated well (positive control) and a representative uninfected, untreated well (negative control). For the negative control sample shown in **Fig. 3a**, the label-free PFU probability map and the final detection mask generated by our system yielded no false-positive detections across all time points, consistent with the corresponding stained ground truth image at 96 hours. PFU formation in positive control samples (HSV-1 infected, ACV-untreated) was similarly monitored over 96 hours, and the resulting label-free PFU probability maps and final detection masks at selected time points of incubation (40 h, 60 h, and 80 h) are presented in **Fig. 3b**. Notably, the computationally predicted PFU locations and plaque areas inferred at these earlier time points exhibit strong spatial concordance with the endpoint stained ground-truth image at 96 hours, firmly validating the system's effectiveness in saving a significant amount of time. An additional representative positive sample tracked across various time points is detailed in **Fig. 4a**. A continuous time-lapse visualization of a representative well, depicting the temporal progression of the probability maps and PFU detection masks alongside the 96-h stained ground truth, is provided in **Supplementary Video 3**.

Our system is also capable of resolving PFU clusters, owing to its early detection and time-lapse monitoring capabilities. As shown in the zoomed-in images in **Fig. 3b**, four distinct PFUs were successfully identified by our system in a region that appeared as a single large plaque in the endpoint-stained ground-truth image. While these individual PFUs are clearly distinguishable in our probability maps and final PFU detection masks, they cannot be accurately counted solely through conventional visual inspection of an endpoint assay. By de-mixing these temporal trajectories, our platform achieves superior analytical precision compared to endpoint assays, which inherently underestimate viral load when distinct infection events merge into a single macroscopically visible plaque. Additional examples of resolved PFU clusters are provided in **Supplementary Figure 2**.

To further quantify the early detection capabilities of our system, blind testing was performed on 3 *new, unseen* ACV-untreated HSV-1-infected wells (positive controls), encompassing a total of 113 PFUs. Notably, the label-free system detected initial PFU formation as early as 28 hours into incubation. The detection rate was defined as the number of system-detected PFUs divided by the total PFU count confirmed via the 96-h stained ground truth (endpoint). Using this metric, our system achieved a detection rate exceeding 80% at 55 hours and 90% at 70 hours (**Fig. 4b**), representing a reduction of ~26 hours compared to the standard 96-h incubation required for manual endpoint inspection. Furthermore, when evaluated on 3 *new, unseen*, uninfected wells with $\sim 2.65\times10^{6}$ distinct 4D test tensors (negative controls), the system produced no false positives at any time point, despite operating in a fully label-free manner.

**Evaluation of our system on HSV-1 infected, ACV-treated samples**

To evaluate the system's utility for antiviral drug screening, we monitored a total of 300 PFUs from 12 HSV-1-infected samples treated with four different ACV concentrations (1.02, 2.04, 3.06, and 5.12 μg/mL; 3 wells per concentration) over 96 hours of incubation; since the training and validation PFU data were never exposed to antiviral drugs, this constitutes a test for external generalization of our system. In a typical PRA experiment with an active antiviral compound, both the PFU count and plaque size are expected to decrease. Consistent with this, our PFU probability maps and final detection masks successfully captured a progressive reduction in PFU numbers and spatial areas with increasing ACV concentrations, in agreement with the stained ground truth images (T = 96 h) shown in **Fig. 5**. The system also demonstrated exceptional sensitivity in identifying highly heterogeneous and severely inhibited PFUs under higher drug concentrations. For instance, at 3.06 μg/mL of ACV, our system successfully resolved small-scale PFUs within a heterogeneous population. Moreover, at the highest drug concentration (5.12 μg/mL), where PFUs were severely stunted and largely indistinct in the 96-h stained ground truth images, our system still detected them clearly. As highlighted in the magnified insets of **Fig. 5**, at 5.12 μg/mL of ACV, two small PFUs (labeled in red circles with yellow arrows, <200 μm in diameter) were detected at 90 hours, which are extremely difficult to resolve through manual human inspection of the stained endpoint ground truth images. Additional examples demonstrating the small-scale PFU detection capability of our system at higher drug concentrations are provided in **Supplementary Figure 3**, further highlighting the sensitivity of the approach.

To illustrate the longitudinal quantification and comprehensive evaluation of the dose-dependent changes in PFU numbers and PFU spatial areas across increasing ACV concentrations, we extracted four metrics from the binary detection masks generated by our system: (1) the total PFU count, $N$; (2) the total PFU area coverage, $A$, defined as the percentage of PFU-positive pixels relative to the total well pixels; (3) the incremental change of PFU count (new PFU formation), $\Delta N$; and (4) the incremental increase in PFU area coverage (PFU area growth), $\Delta A$, as shown in **Fig. 6**. The cumulative metrics ($N$ and $A$) are evaluated at every detection time point with an hour interval during the incubation, whereas the incremental metrics ($\Delta N$ and $\Delta A$) are calculated over specific temporal windows (an initial 0–24 hours period, followed by subsequent 12-h intervals, as shown in **Fig. 6a-b**). We evaluated these four quantitative metrics at four ACV concentrations (1.02, 2.04, 3.06, and 5.12 μg/mL) and an ACV-untreated condition serving as the positive control, with all samples infected with the same HSV-1 inoculum. Each condition included three replicates (n = 3), with each curve and bar in **Fig. 6** representing the mean ± standard deviation, respectively. A detailed analysis of the drug response dynamics under these conditions is elaborated below.

In the untreated samples ("no drug" condition), HSV-1 plaque dynamics followed a characteristic pattern (see **Fig. 6a**). No plaques were detected during the first 24 hours post-infection, which is consistent with the eclipse phase of viral replication[50]. New plaque formation then accelerated between 24-36 hours ($\Delta N \approx 6$), peaked sharply at 36-48 hours ($\Delta N \approx 19$), and then decreased as the number of susceptible host cells in the monolayer became gradually limited. Furthermore, the dynamic pattern reflected in $\Delta N$ is strongly corroborated by the trends observed in $\Delta A$, as shown in **Fig. 6b**. Specifically, PFU spatial expansion proceeded rapidly at 36-60 hours, with the average $\Delta A$ over 8% at 36-48 hours and 7% at 48-60 hours. This PFU area coverage growth $\Delta A$ maintained increments of 3–5% through the later 12-h intervals, reflecting ongoing cytopathic spread from already-formed plaques. The temporal progression of the cumulative metrics ($N$ and $A$) is further detailed in **Fig. 6 c-d**.

ACV treatment disrupted this typical progression of HSV-1 infection in a concentration-dependent manner. At an ACV concentration of 1.02 μg/mL, new plaque formation was delayed, with the peak of $\Delta N$ occurring at 48-60 hours, compared to 36-48 hours in the untreated condition. The peak value itself ($\Delta N \approx 13$) was lower than the untreated baseline ($\Delta N \approx 19$). Moreover, the PFU area coverage growth ($\Delta A$) followed a similar delayed and suppressed pattern at this ACV concentration. Specifically, the peak of $\Delta A$ was shifted to 60-72 hours, contrasting with the 36-60 hours observed in the untreated condition, and the peak $\Delta A$ value itself is also lower than the untreated baseline (see **Fig. 6a-b**). At the end of the incubation period (96 hours), the total PFU count ($N$) at 1.02 μg/mL of ACV eventually reached a level *comparable* to the untreated condition, as demonstrated in **Fig. 6c**, which was driven by a second peak in new plaque formation between 84-96 hours. This secondary wave of virus replication under 1.02 μg/mL of ACV revealed a heterogeneous mixture of large and small PFUs in the 96-hour-stained ground truth (**Fig. 5**), representing two distinct temporal phases of viral progression. Our platform's ability to resolve this secondary replication wave provides unique insights into drug-exhaustion kinetics and viral population heterogeneity, capturing pharmacodynamic shifts that are fundamentally obscured by endpoint-only measurements. The inhibitory effect of the drug is also clearly evident in the total PFU area coverage $A$. This indicates that although some viral replications may resume at later stages, the drug still substantially limits the overall extent of plaques. As shown in **Fig. 6d**, $A$ only reached ~15-20% under 1.02 μg/mL of ACV, compared to ~30% in untreated conditions.

These findings underscore an important advantage of our continuous PFU monitoring platform: rather than relying on potentially misleading 96-h endpoint counts, the temporal shift in the kinetic peaks of $\Delta N$ and $\Delta A$ allows for the conclusive determination of drug effectiveness as early as ~60 hours by cross-confirming the peak-shift information. This reveals that antiviral effects manifest as temporally distinct modulation of plaque initiation and expansion, which cannot be inferred from endpoint PFU counts alone.

In the presence of 2.04 μg/mL of ACV, all the metrics were further suppressed across 96 hours of incubation: the peak $\Delta N$ dropped to ~10 and was delayed to 60-72 hours; $N$ and $A$ plateaued at ~25-30 PFUs and ~10–15% coverage, respectively. At 3.06 μg/mL of ACV, new plaque formation was tightly restricted, with $\Delta N$ smaller than ~7 across three consecutive time intervals within the 60-96 h window; correspondingly, the total PFU count $N$ gradually plateaued at ~20-23 by 96 hours (**Fig. 6 c-d**). At the highest ACV dose (5.12 μg/mL), the detection results generated by our system faithfully capture the severely restricted viral progression expected under such strong drug inhibition. The HSV-1 replication was almost completely suppressed at this highest concentration, with $\Delta N$ (mean of 3 replicas) remaining <3 across all 12-h intervals throughout 96 hours of incubation. Similarly, PFU area coverage growth $\Delta A$ stayed consistently below 1%, resulting in the final total PFU area coverage ($A$) achieved at 96 hours being <3%, indicating near-total inhibition of both new plaque formation and spatial plaque expansion at 5.12 μg/mL of ACV.

We also observed that replicate variability (n = 3), represented by the shaded regions for $N$ and $A$ and error bars for $\Delta N$ and $\Delta A$ in **Fig. 6**, was highest in the untreated wells and at lower ACV concentrations (e.g., 1.02 and 2.04 μg/mL). This variation reflects stochastic differences in the initial viral infection and early replication events, whereas higher concentrations produced highly reproducible, low-variability outputs. In the late phase of incubation (≥72 hours), the inhibitory effect of ACV on plaque expansion weakened at lower doses. As discussed above, this weakened suppression allowed for a secondary wave of PFU growth, resulting in a morphologically heterogeneous population of plaques; see **Fig. 5**. Only the 5.12 μg/mL of ACV condition maintained clear separation with other dose concentrations throughout the entire incubation period, suggesting a concentration threshold required for sustained suppression of HSV-1 plaque progression.

These dynamic quantitative data profiles continuously obtained by our system demonstrate that ACV reduces the total number of HSV-1 plaques in a dose-dependent manner, effectively delays and lowers the peak of new PFU formation, and restricts the spatial expansion of plaque area. Crucially, our automated, label-free platform extracts these distinct, time-resolved antiviral effects without the destructive endpoint processing based on staining. This continuous tracking enables rapid evaluation through direct observation of data progression; for instance, by analyzing the temporal shifts in new PFU formation peaks, drug efficacy can be determined within ~60 hours, saving significant time. Importantly, at higher drug concentrations (e.g., 3.06 μg/mL and 5.12 μg/mL of ACV), this detection timeframe can be further reduced. By providing substantially richer viral susceptibility data than conventional PRA, together with the capability to resolve small-scale PFUs and PFU clusters, our platform improves quantitative PFU analysis, detection accuracy and the time efficiency of dose–response characterization, highlighting its promising translational potential for dynamic antiviral resistance monitoring for various viral pathogens.

## DISCUSSION

Our automated, rapid, and label-free PRA platform demonstrates a transformative approach to antiviral susceptibility testing by continuously tracking the spatio-temporal dynamics of viral infection. Traditional endpoint assays are limited by their reliance on chemical fixation and staining, which destroy the cell monolayer and terminate viral progression, thereby making continuous observations impossible. In our framework, antiviral susceptibility is recast as a dynamic label-free phenotyping problem, where infection trajectories—not just endpoints—define drug response.

While prior studies have applied machine learning and label-free imaging to detect viral plaques or cytopathic effects, these approaches remain focused on static detection or classification tasks. In contrast, our work introduces a fully time-resolved analytical framework that enables continuous quantification of antiviral response dynamics, bridging a critical gap between detection and functional susceptibility testing. Operating non-destructively, our label-free platform successfully quantifies dose-dependent antiviral responses and exploits unique spatio-temporal features learned by the PFU detection neural network to identify the formation of early-stage, small-scale PFUs before they become visible to the human eye. Crucially, this dynamic monitoring capability enables the precise enumeration of dense PFU clusters that would otherwise coalesce into indistinguishable single plaques in standard 96-h stained ground truth images. Ultimately, this platform eliminates the need for hazardous staining reagents and for subjective, laborious manual inspection, yielding a substantially richer data profile for evaluating dynamic antiviral efficacy, which can be seamlessly used to extract other standard pharmacological metrics. More broadly, this work establishes a conceptual shift from endpoint-based assays to continuous label-free measurement frameworks for virology and drug discovery.

We should note that the outstanding performance and sensitivity of our system are enabled by our TFT image sensor with an ultra-large active area of 100 $cm^2$, which corresponds to an imaging area of ~15.4 $cm^2$ at the sample plane. Using a single-shot image acquisition, the platform achieves ultra-high throughput without mechanical scanning of the sample well. This single-shot image acquisition paradigm critically simplifies the computational pipeline: it bypasses the need for complex image-stitching algorithms to reconstruct the full well image and inherently guarantees precise spatial alignment across time points during the incubation period. Consequently, there is no requirement for computationally expensive elastic image registration algorithms[51,52] to correct for mechanical drift. This spatio-temporal alignment is the cornerstone of our label-free differential processing, enabling remarkably clean subtraction of static cellular backgrounds, eliminating false positives. By isolating the active cytopathic signatures in these differential images, our deep learning algorithm also achieves a drastic increase in detection sensitivity compared to relying solely on raw intensity data (see **Supplementary Fig. 1**), thereby unlocking the system's capacity for early PFU identification. Furthermore, TFT image sensors are inherently cost-effective and can be manufactured even more economically via roll-to-roll mass production. Critically, the low power consumption of TFT image sensors results in minimal heat dissipation, ensuring a stable thermal environment within the incubator; this prevents hardware-induced perturbations to viral replication kinetics or host-cell metabolism during multi-day longitudinal monitoring—a frequent limitation of Complementary Metal-Oxide-Semiconductor (CMOS) or Charge-Coupled Device (CCD)-based imaging systems.

To determine the optimal input features for our PFU detection neural network, we conducted a comprehensive channel selection ablation study, as detailed in **Supplementary Figure 1**. Visually, the raw intensity image, reconstructed amplitude, and phase channels are dominated by static cellular background noise, which severely buries the cytopathic signatures of expanding PFUs (**Supplementary Figure 1a**). Quantitatively, the network models trained solely on these static channels exhibited poor validation sensitivity, such as 62.6% for the raw image and 0% for the reconstructed amplitude, as shown in **Supplementary Figure 1b**. In contrast, the differential channels effectively subtract the static background, isolating the dynamic spatio-temporal changes caused by viral replication. Consequently, the Differential Channel 2 ($\Delta T = 12$ h) alone achieved a substantially higher validation sensitivity of 83.0%. Furthermore, our double-channel ablation experiments (reported in **Supplementary Figure 1c**) demonstrated that combining Differential Channel 1 ($\Delta T = 8$ h) and Differential Channel 2 ($\Delta T = 12$ h) yielded the best overall performance, achieving a validation sensitivity of 83.5% and a validation specificity of 99.3%. Notably, appending raw static channels (intensity, amplitude, or phase) to the differential inputs did not improve sensitivity and, in some cases, degraded specificity by reintroducing background noise. Therefore, the combination of two differential channels (1 and 2) was selected as the optimal input for our PFU detection neural network.

Furthermore, the inclusion of the temporal dimension ($D = 8$), i.e., two differential images with different subtraction intervals ($\Delta T = 8$ and 12 h) across 8 consecutive frames, is critical to the success of our PFU detection strategy. While differential maps inherently capture dynamic PFU information by representing change (or “growth speed”), the differential image at a single frame remains ambiguous, as it cannot clearly distinguish between transient environmental disturbances and sustained biological growth. For instance, artifacts such as the sudden deposition of a dust particle or the rupture or transient expansion of a micro-bubble can generate a strong, high-contrast signal in differential maps at a single time point, mimicking the appearance of a PFU start. However, these events typically represent non-sustained changes. By stacking these differential maps into a sequence, the network captures the higher-order spatio-temporal evolution of these dynamics, effectively filtering out transient artifacts and specifically isolating the continuous, radial propagation characteristic of true PFUs.

## MATERIALS and METHODS

### Safety

All cell culture and virus culture procedures were performed under Biosafety Level 2 (BSL-2) containment and in accordance with the UCLA Environmental, Health and Safety (EH&S) rules and regulations. All experiments were conducted in a certified Class II biological safety cabinet, employing aseptic techniques throughout.

### Studied organisms and their storage

Herpes simplex virus type 1 (HSV-1; ATCC VR-260) and Vero E6 cells (Vero C1008, clone E6; ATCC CRL-1586), an African green monkey kidney epithelial cell line, were obtained from the American Type Culture Collection (ATCC). Upon arrival, both virus and cell stocks were stored in liquid nitrogen in accordance with ATCC protocols.

### Preparation of solutions

*Complete medium I (cell culture)*: Eagle's Minimum Essential Medium (EMEM; cat. no. 30-2003, ATCC) served as the basal medium for cell culture preparation. Complete growth medium was prepared by supplementing EMEM with 10% (v/v) fetal bovine serum (FBS; cat. no. 30-2021, ATCC).

*Complete medium II (overlay medium):* This medium was prepared by supplementing EMEM with 2.45% (v/v) FBS to be used in preparing the layering solution.

*Agarose stock solution (4.0% w/v):* The stock of 4.0% (w/v) agarose (cat.no. MP11AGR0050, Fisher Scientific) was prepared in reagent-grade water (cat.no. 23-249-581, Fisher Scientific) and mixed thoroughly in a glass bottle until dissolved. The solution was sterilized by autoclaving at 121 °C for 15 minutes, aliquoted into 2.0 mL portions, and stored at 4 °C until use.

*Agarose overlay solution with Acyclovir (ACV):* A 4.0 mL aliquot of 4.0% agarose stock was melted in a microwave for approximately 40 seconds and then equilibrated in a 70°C water bath. In parallel, complete medium II was warmed in the same water bath for 5 minutes. Both were then transferred to a biosafety cabinet, where 19.0 mL of prewarmed complete medium II was mixed with 1.0 mL of the melted agarose stock to prepare the overlay base solution. Later, acyclovir (ACV; cat. no. A4669, Sigma-Aldrich) stock solution was prepared at a concentration of 23 µg/mL in 1:1 dimethyl sulfoxide: water (DMSO; cat. no. D8418, Sigma-Aldrich) and stored at −20°C. From this ACV stock, agarose overlay solutions containing ACV at final concentrations of 1.02, 2.04, 3.06, and 5.12 µg/mL were prepared with the overlay solution maintained at 45–50°C during preparation.

**Cell propagation, cell subculturing, and well-plate preparation**

For cell propagation, Vero cells were taken from the liquid nitrogen tank, and, after thawing, 0.2 mL of cells was mixed with 9.8 mL of prewarmed complete medium I and cultured in a T-75 flask (cat. no. FB012937, Fisher Scientific). The cell flask was maintained at 37 °C in a humidified incubator with 5% $CO_2$, and cell adherence to the flask was observed every other day with a phase-contrast microscope until the cell monolayer reached ~95% confluency.

Prior to cell subculturing, all liquid reagents, trypsin–EDTA (cat. no. 30-2101, ATCC), EMEM, D-PBS (cat. no. 20012-027, Gibco), and FBS were warmed to 37°C to maintain physiological conditions. A confluent T-75 flask was removed from the incubator and transferred to the biosafety cabinet. The cell monolayer was gently rinsed once with 10.0 mL of PBS (pH 7.2) and applied along the flask wall to minimize shear forces. After PBS was aspirated, 4.0 mL of prewarmed trypsin–EDTA was added, and the flask was gently shaken to ensure uniform coverage of the reagent over the cells and effective trypsinization. After incubation for 5 minutes, the flask was gently tapped and examined under dark-field microscopy to confirm the detachment by the presence of rounded, freely suspended cells. Upon detachment of the cells, 4.0 mL of prewarmed EMEM was added to the flask and was mixed by pipetting across the flask surface using a serological pipette to ensure complete dissociation of the monolayer. Then, the cell suspension was transferred to a sterile 15.0 mL conical tube and centrifuged at 300 x g for 5 minutes in a balanced rotor (ComboXL, LW Scientific). After centrifugation, the supernatant was carefully aspirated without disturbing the cell pellet. The cells were resuspended in 4.0 mL of complete medium I and gently dispersed to obtain a uniform single-cell suspension. An aliquot was mixed 1:1 with trypan blue (cat. no. SV30084, HyClone), and cell density and viability were determined using a CelloMeter automated cell counter (Nexcelom Biosciences).

Before each cell subculture or well-plate preparation, the cell suspension should be gently mixed to prevent aggregation. For establishing a new T-75 flask, cells were diluted to a final concentration of $1.0x10^4$ cells/mL in 10 mL of complete medium I. For a 6-well plate preparation, the cell count was set to $2.0x10^5$ with 10.0 mL of complete medium I and 2.0 mL of this cell suspension dispensed into each well. Then, the flask and the 6-well plate were placed in the $CO_2$ incubator at 37 °C for 1 week and 2 days, respectively.

**Virus propagation**

HSV-1 was propagated in a T-75 flask with Vero cells, with a cell monolayer reaching at least 90% confluency. For this purpose, FBS, PBS, and EMEM were prewarmed to 37 °C before propagation. Once the flask was ready, it was rinsed with 10 mL of D-PBS, and then 6.0 mL of EMEM was added. Then, the flask was infected with 20 µL of HSV-1 stock at a multiplicity of infection (MOI) of 0.07 and incubated in 5% $CO_2$ at 37 °C for 72-96 h. The appearance of rounded, detached cells confirmed successful propagation. The solution was transferred to a 50.0 mL conical tube (cat. no. 06-443-20, Fisher Scientific) and tightly sealed with parafilm prior to centrifugation at ~2,600 g for 10 minutes using a centrifuge with swing-out rotors. The supernatant containing the virus was collected and transferred to a new tube. Then, the virus suspension was aliquoted into 1.0 mL cryogenic vials with O-ring (cat. no. 5000-1012, Fisher Scientific) and stored in a liquid nitrogen tank.

**Plaque reduction assay (PRA) with HSV-1**

Cells in each well (6-well plate) were infected with 100 µL of diluted HSV-1 suspension (dilution factor of $0.8–1×10^{-5}$ from a stock of approximately $9.0×10^7$ PFU/mL). The plate was then incubated for 1 hour with gentle rocking at 15-minutes intervals to facilitate viral adsorption. Following incubation, the inoculum was aspirated, and 2.0 mL of agarose overlay solution prepared with the target acyclovir concentration (1.02, 2.04, 3.06, or 5.12 µg/mL) was applied to each well at 30–35°C. After solidification at room temperature for 40 minutes, the plate was placed onto a TFT-based lens-free imaging system housed within the incubator and monitored for 96 hours at 1-hour intervals under standard culture conditions (37°C, 5% $CO_2$).

Following incubation, the well plate was transferred to the biosafety cabinet and allowed to equilibrate for 15 minutes. A fixation solution was prepared by diluting 16% paraformaldehyde (cat. no. 15710-S, Electron Microscopy Sciences) to a final concentration of 6.25% in methanol, and 1.0 mL of this solution was added to each well. The plate was then incubated for 50 minutes to inactivate the virus and fix the cells. The soft agar layer was then gently removed, and 1.0 mL of crystal violet solution (1:1 crystal violet: water; cat. no. 94448, Sigma-Aldrich) was added to each well. The plate was placed on an orbital shaker (Benchmark, Incu-Shaker 10L) at 70 rpm for 10 minutes, after which the crystal violet solution was removed, and excess dye was washed away with distilled water. The plate was then left to dry in a fume hood until further image analysis.

**TFT-based lens-free imaging setup**

The compact lens-free PFU imaging system comprises: (1) a coherent illumination source, (2) a sample holder, (3) a large-area TFT-based custom-designed image sensor, (4) a customized printed circuit board (PCB) for control, and (5) controlling software. For illumination, we used a green laser diode ($\lambda$ = 515 nm, 2 nm bandwidth, 0.17 mm emission diameter, Osram PLT5510). The sample holder was positioned at a distance of $z_1$ = 140 mm below the illumination source, and the image sensor was placed at $z_2$ = 217 mm beneath the sample, resulting in a fringe magnification factor[45] ($F = (z_1 + z_2) / z_1$) of 2.55. Meanwhile, the center of the laser source, the center of the observation well, and the center of the sensor active area were aligned along the optical axis. This configuration allows the diverged laser beam to fully cover the detection well while ensuring the magnified lens-free diffraction patterns of samples fit within the sensor's active area with sufficient marginal redundancy. Image acquisition was performed using a custom-designed ultra-large-area TFT image sensor (Japan Display Inc.), featuring a 10 cm × 10 cm active area (1000 × 1000 pixels, pixel size: 100 μm). The chosen fringe magnification factor (2.55×) effectively mitigates the resolution constraints imposed by the sensor's large pixel size, ensuring that even early-stage PFU features are sufficiently enlarged to span multiple pixels and support effective deep learning-based neural network identification. Illumination control was implemented via a custom PCB integrating an Arduino Micro microcontroller and a constant-current LED driver (TLC5916, Texas

Instruments). A custom-designed software interface (**Supplementary Figure 4**) manages the time-lapse imaging sequences by synchronizing the illumination triggering with the TFT sensor readout.

**Time-lapse image data acquisition**

HSV-1 samples were loaded into the lens-free imaging system to start 96-hour time-lapse imaging with a 1-hour acquisition interval. To minimize light exposure and prevent potential perturbations to viral infection dynamics, the laser diode remained inactive between imaging cycles. During each acquisition event, the laser diode was triggered to provide stable illumination encompassing the 110 ms exposure/integration time of the TFT image sensor. A dynamic representation of the time-lapse imaging process captured by our device is shown in **Supplementary Video 1**. Computational analysis was initiated at the 22$^{nd}$ hour of incubation, after which the acquired images were processed by our neural network-based algorithm for automated PFU detection and quantification. To establish comparative ground truths against our system's outputs, the samples were fixed, chemically stained, and manually counted after 96 hours of incubation following the traditional PRA workflow.

**Image pre-processing and differential analysis**

The raw time-lapse images acquired by the lens-free TFT-based imaging system serve as the inputs to our computational PFU detection pipeline. Prior to neural network training/validation or testing, these image sequences were subjected to a pre-processing protocol that served two objectives. First, illumination normalization and spatial registration were applied to compensate for fluctuations in light source intensity and potential positional shifts induced by environmental vibrations (e.g., from incubator operation). Second, to optimize the data for effective network learning, a differential analysis scheme was introduced to suppress static background and specifically amplify the dynamic contrast of PFU features (see **Supplementary Figure 1** for a visualization of this contrast enhancement). The implementation details of these three sequential steps are elaborated below.

First, illumination normalization is applied to establish a consistent image-intensity baseline across the temporal sequences. For each frame T in the sequence, a static, feature-free background region $R_b$ (200 × 200 pixels) was identified to monitor intensity variations. The global intensity of the current raw frame $I_{raw}^{(T)}$ was scaled to match the illumination level of the initial reference frame (T = 0 h). The corrected image $I_{corr}^{(T)}$ is computed as:

$$I_{corr}^{(T)} = I_{raw}^{(T)} \cdot \frac{\mu_{R_b}^{(0)}}{\mu_{R_b}^{(T)}} \quad (1),$$

where $\mu_{R_b}^{(T)}$ and $\mu_{R_b}^{(0)}$ denote the mean intensity values of the background region $R_b$ in the current frame (T) and the initial frame, respectively. This linear scaling effectively compensates for global brightness drift, ensuring temporal uniformity throughout the observation period.

With illumination consistency established, the pre-processing pipeline subsequently addresses potential spatial misalignments introduced by environmental vibrations. Unlike scanning-based imaging systems, our large field-of-view (FOV) single-shot-based TFT architecture is inherently free from local geometric distortions or stitching artifacts. Consequently, the registration process requires only rigid global alignment to compensate for macroscopic positional shifts. We employed a cross-correlation-based algorithm to compute the optimal translational displacement vector $(\Delta x,\ \Delta y)$ between consecutive frames. Based on the peak correlation response, the current frame was spatially shifted to align strictly with the previous frame, and subsequently center-cropped to eliminate boundary artifacts. Then, a dual-channel temporal differencing strategy was employed to highlight dynamic PFU growth from the static cellular background. The differential map $I_{diff}^{T}$ at frame T is calculated by subtracting the average intensity of a preceding historical temporal window (offset by $\Delta T$) from that of the current window:

$$I_{diff}^{(T)} = \frac{1}{M}\sum_{i=1}^{M} I_{corr}^{(T-i)} - \frac{1}{M}\sum_{j=1}^{M} I_{corr}^{(T-j-\Delta T)} \quad (2),$$

where M represents the average window size (set to 3 frames to suppress high-frequency noise). The temporal gap parameter, $\Delta T$, was configured with two distinct intervals: $\Delta T$ = 8 h and $\Delta T$ = 12 h, generating two differential channels, $I_{diff_1}^{(T)}$ and $I_{diff_2}^{(T)}$, respectively. This operation effectively cancels out the static background of the uninfected cell monolayer while retaining the dynamic features generated by PFU expansion. The resulting difference maps were subsequently processed with a median filter and a Gaussian low-pass filter to eliminate residual pixel-level noise, yielding high-contrast inputs specifically optimized for the downstream PFU detection network.

**Preparation of the training and validation datasets**

Leveraging the pre-processed differential channels, $I_{diff}^{(T)}$, generated in the previous steps, we constructed a comprehensive spatio-temporal dataset to train and validate the PFU detection neural network. For positive training/validation instances, we manually identified PFU centroids on the differential images of infected wells at T = 96 h, using the corresponding chemically stained images as ground truth. Since the differential image stacks were already spatially registered with high precision, the terminal PFU centroid coordinates were applied to the preceding time frames to consistently localize the viral plaques. This retrospective strategy effectively circumvents the ambiguity of manually pinpointing early-stage, low-contrast infection events. For negative training/validation instances, coordinate selection followed a strict exclusion protocol to ensure validity: within infected wells, centroids were manually placed to avoid PFU regions, whereas in uninfected negative control wells, centroid coordinates were generated via random sampling.

Centered on these spatially verified coordinates, $I_{diff}^{(T)}$ stacks were extracted to a 4D spatio-temporal tensor, denoted as $X \in \mathbb{R}^{C \times D \times H \times W}$, which serves as the network input. Here, $C = 2$ corresponds to the two differential feature channels ($I_{diff_1}^{T}$ and $I_{diff_2}^{T}$), $D = 8$ represents the consecutive time points, and H and W denote the spatial height and width of the 50 × 50-pixel crop. For every positive (PFU expansion) and negative (non-PFU) sample defined by these centroids, the training/validation videos were augmented over time by extracting 8-frame sequences with a 1-hour stride across the full time-lapse series, effectively multiplying the volume of available training/validation data.

Although the dataset curation protocols described thus far establish a robust baseline, relying exclusively on manual selection and random sampling of negative samples introduces an inherent selection bias: these methods often fail to capture rare, high-resemblance artifacts that statistically mimic PFU features. To overcome this limitation and strictly suppress false positives, we expanded the negative dataset using an iterative hard-negative mining strategy. An initial classifier trained on the baseline data was deployed to scan the full FOV of uninfected control wells. Any 'PFU candidates' detected in these PFU-free wells were identified as false positives and then integrated into the negative training set for the subsequent iteration. This training and mining cycle was repeated four times until the model generated no new false positives in uninfected control wells. Notably, this hard-negative mining strategy was applied exclusively to the training set of uninfected control wells; the validation and blind-testing sets remained strictly isolated and untouched throughout the entire process. This data-driven hard-negative mining strategy forces the training dataset to include ambiguous, challenging examples (specifically those that can deceive the PFU model), ultimately achieving a zero false-positive rate—a level of robustness unattainable through standard random sampling or manual data selection.

The final dataset comprises one positive category and three distinct negative categories: background negatives (manually labeled from infected wells), control negatives (randomly selected from uninfected control wells), and hard negatives (obtained via a hard-negative mining strategy). The detailed composition is summarized in **Table 1**.

**Table 1. Training and Validation Dataset Overview.**

| Data Class | Data Source | Description | Training ($N_{sample}$ / $N_{exp}$)* | Validation ($N_{sample}$ / $N_{exp}$)* |
|---|---|---|---|---|
| Positive (PFU) | Infected wells | Center-cropped sequences of confirmed viral plaques | 13,267 / 19 | 5,051/ 5 |
| Background Negative | Infected wells | Manually selected from verified PFU-free regions | 14,384 / 19 | 8,907 / 5 |
| Control Negative | Control wells | Randomly crops from uninfected wells | 6,389 / 5 | 710 / 5 |
| Hard Negative | Control Wells | False positives (artifacts) identified via iterative mining | 23,256 / 5 | N/A |

* Note: Data counts are presented as: # of training/validation videos (4-D tensors) / # of independent experiments.

**Architecture of the PFU detection network and training schedule**

The PFU detection network is built on a Dense-Net backbone[48], with the 2D convolutional layers replaced by Pseudo-3D[49] (P3D) building blocks. The overall architecture is shown in **Fig. 7**. As described in the previous section, the network input is a 4D tensor of size 2×8×50×50 (differential channels × time × height × width). The network begins with a 3D convolutional layer (32 output channels, kernel size of 1×7×7, and stride of 1×1×1), which extracts spatial features across each time frame while preserving temporal resolution. This is followed by a 3D max-pooling layer (kernel size of 1×3×3, stride of 1×2×2), which reduces the spatial resolution while maintaining temporal continuity, yielding a feature map of size 32×8×25×25. Subsequently, the architecture employs three 3D dense blocks interspersed with 3D transition blocks. Feature maps within each dense block are densely connected, where the outputs from all preceding layers are concatenated along the channel dimension. Each dense layer alternately employs three distinct types of P3D layers (Type A, B, and C), as detailed in **Fig. 7**. Between these dense blocks, 3D transition blocks are inserted to compress the feature dimensions progressively. After the third 3D transition block, the temporal dimension of the feature tensor is reduced to 1, resulting in a 103×1×3×3 feature map. Subsequent processing uses a 2D dense block comprising three 2D dense layers, each consisting of a sequence of 1×1 and 3×3 convolutions with dense connectivity. Finally, a global 2D average pooling layer (kernel size of 3×3) reduces the feature map into a 1D vector. A fully connected layer then converts this vector into the probability of the input 4D tensor belonging to the PFU or non-PFU class. The network was trained using the weighted cross-entropy loss function, which is defined as:

$$\mathcal{L}(p,g) = \sum_{k=1}^{K} \sum_{c=1}^{2} -w_c g_{k,c} \log\left(\frac{\exp(p_{k,c})}{\sum_{c=1}^{2} \exp(p_{k,c})}\right) \qquad (3),$$

where $p$ denotes the unnormalized score of each class (i.e., PFU or non-PFU) before the SoftMax layer, $g_{k,c}$ represents the corresponding one-hot encoded ground-truth label (where $g_{k,c} = 1$ if the instance belongs to class $c$, and 0 otherwise), $K$ is the total number of training instances in one batch, and $w_c$ is the class-specific weight applied to address class imbalance. In our case, $w_c$ is defined as $w_c = 1 - d$, where $d$ is the percentage of the samples in one class ($d$ = 23.16% for the positive class and $d$ = 76.84% for the negative class). The model was trained using the Adam optimizer with an initial learning rate of $1\times10^{-4}$ and momentum parameters $\beta_1 = 0.9$ and $\beta_2 = 0.999$. A step-based learning rate scheduler (StepLR) was employed, with the learning rate multiplied by 0.7 every 30 epochs. The final model was selected based on the best accuracy on the validation dataset.

**Image processing for blind testing**

The post-processing pipeline was designed to transform local network probability predictions into global quantitative metrics of viral progression, as demonstrated in **Supplementary Figure 5**. The generation of quantitative detection results commenced at T = 22 h. This specific start time was set due to the necessary data accumulation buffers: the initial 15 hours were required to generate the first valid differential feature maps ($I_{\text{diff_1}}^{\text{T}}$ and $I_{\text{diff_2}}^{\text{T}}$), followed by an additional 7-hour window to construct the first complete 8-frame spatio-temporal sequence for network inference. Upon satisfying this temporal buffer, the network processed the full sample FOV using an overlapping sliding-window strategy, as shown in **Supplementary Video 2**. A 50 × 50-pixel processing window traversed the entire 1000 × 1000-pixel FOV with a stride of 10 pixels, yielding a 96 × 96 probability grid $P_{raw}^{(T)}$ for each time frame T (**Supplementary Figure 5a**), effectively mapping locally inferred probability scores to global spatial coordinates. To ensure analysis was restricted to the valid optical field, a circular region of interest (ROI) was applied to this probability grid. The ROI was defined by a radius of $r$ = 40 grid pixels.

Following the spatial reconstruction, $P_{raw}^{(\text{T})}$ was binarized using a decision threshold of 0.5 to generate initial binary masks $D_{raw}^{(\text{T})}$ (**Supplementary Figure 5b**), where the pixels with probabilities ≥ 0.5 are classified as PFU regions (positive) and those below the threshold as non-PFU background (negative). To mitigate residual errors and strictly reject non-PFU artifacts, a morphological filter was applied to $D_{raw}^{(\text{T})}$. Connected components were analyzed based on geometric properties; objects were discarded if their area fell below a minimum pixel threshold ($s_{min}$= 5 pixels) or if their circularity dropped below 0.7. This step yields the cleaned binary mask output $D_{clean}^{(\text{T})}$ (**Supplementary Figure 5c)**, in which only valid, plaque-like structures are retained for a given time frame T.

Following the morphological filtration, we computed a masked probability map $P_{clean}^{(\text{T})}$ (**Supplementary Figure 5d**) for each time frame T:

$$P_{clean}^{(\mathrm{T})}(x, y) = P_{raw}^{(\mathrm{T})}(x, y) * D_{clean}^{(\mathrm{T})}(x, y) \quad (4),$$

where $x$ and $y$ are spatial coordinate indices and $*$ corresponds to pixel-wise multiplication. This operation effectively zeroes out background noise and artifact regions while retaining the probability distribution within valid plaque regions. Subsequently, a probability map $P_{proj}^{(\mathrm{T})}$ (**Supplementary Figure 5f**) was computed to comply with the biological constraint that viral plaque formation is permanent and irreversible. Specifically, the value of each pixel at the current time T was defined as the maximum masked probability observed at that coordinate from the onset of observation (T = 22 h) up to T:

$$P_{proj}^{(\mathrm{T})}(x, y) = \max_{22 \le t \le \mathrm{T}} \{P_{clean}^{(t)}(x, y)\} \quad (5).$$

This *temporal maximum projection* (**Supplementary Figure 5e**) ensures that once a pixel is identified as part of a valid PFU with high confidence, its state is preserved in the longitudinal record, thereby stabilizing the detection against transient signal fluctuations.

Then, we addressed the geometric projection distortions inherent to the lens-free point-source illumination setup. Due to the divergent nature of the illumination wavefront, light rays intersect the sample at increasingly oblique angles towards the periphery of the FOV. This obliquity results in a non-uniform projection onto the flat sensor surface: while magnification in the tangential direction remains relatively consistent, the radial dimension is significantly elongated due to the projection of the inclined light cone onto the planar sensor. Consequently, PFUs, which are morphologically circular in the sample plane, appear as radially stretched ellipses at large field angles. To correct this distortion, a computational correction was applied to $P_{proj}^{(\mathrm{T})}$ using a pre-calibrated geometric transformation model to obtain $P_{final}^{(\mathrm{T})}$ (**Supplementary Figure 5g**), thereby restoring correct plaque morphology across the entire sample FOV. Then, the final detection mask $D_{final}^{(\mathrm{T})}$ (**Supplementary Figure 5h**) was generated by applying a decision threshold of 0.5 to $P_{final}^{(\mathrm{T})}$.

Finally, we developed an automated PFU count algorithm. As infection progresses, new plaques often emerge adjacent to or merge with existing lysed regions, forming a single, connected component in the binary mask. To accurately distinguish new infection events from the existing footprint, we implemented a temporal morphological subtraction algorithm. For each time frame T, we compared the current binary mask with the mask from the previous frame T-1. The incremental growth regions were identified by subtracting the historical footprint from the current detection mask:

$$D_{growth}^{(\mathrm{T})} = D_{final}^{(\mathrm{T})} - D_{final}^{(\mathrm{T}-1)} \quad (6).$$

The final PFU count, $N$, was then defined as the temporal summation of all valid, independent growth events observed throughout the experiment:

$$N = \sum_{t=22}^{\mathrm{T}} \sum_{k=1}^{K_T} \mathrm{One}(Area(C_k^t) \geq s_{min}) \quad (7),$$

where $C_k^t$ denotes the $k$-th connected component within the incremental mask $D_{growth}^{(T)}$, $K_\mathrm{T}$ is the total number of components at time T, $s_{min}$ is the minimum spatial threshold (5 pixels), and $\mathrm{One}(\cdot)$ is the indicator function, which equals 1 if the condition is met and 0 otherwise. This formulation resolves merged plaques by counting only substantial, spatially isolated growth regions as new PFU formation events. In contrast, incremental increases that are contiguous with pre-existing plaques—typically smaller than $s_{min}$ under our hour-to-hour imaging setting—are interpreted as normal plaque expansion and are not counted as additional PFUs.

**Other implementation details**

The image pre-processing and post-network processing workflows were all implemented in MATLAB R2022b (MathWorks, Natick, MA). The code for constructing and training neural network models was written in Python 3.9.12 and PyTorch 1.11.0. The training was executed using a desktop computer equipped with an Intel Core i9-10920X CPU, 256GB of memory, and an Nvidia GeForce RTX 3090 GPU.

## FIGURES

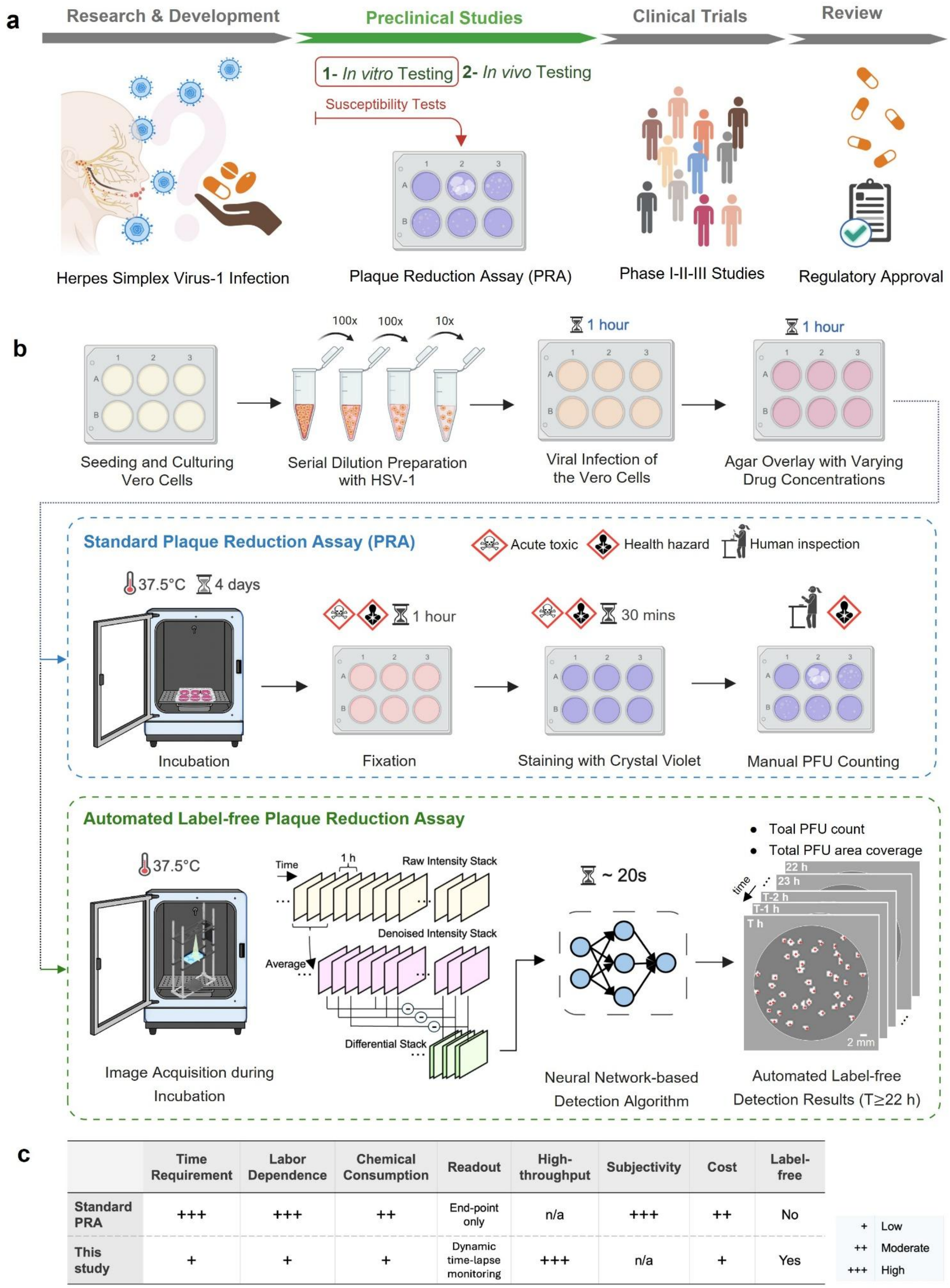

| | Time Requirement | Labor Dependence | Chemical Consumption | Readout | High-throughput | Subjectivity | Cost | Label-free |
|---|---|---|---|---|---|---|---|---|
| Standard PRA | +++ | +++ | ++ | End-point only | n/a | +++ | ++ | No |
| This study | + | + | + | Dynamic time-lapse monitoring | +++ | n/a | + | Yes |

**Figure 1. Overview of the label-free, rapid, and automated Plaque Reduction Assay (PRA) platform for antiviral susceptibility testing. (a)** Contextualization of the *in vitro* PRA within the preclinical stage

of the antiviral drug development pipeline. **(b)** Workflow comparison between the standard PRA and the proposed automated label-free system. Following the initial cell culturing and viral infection steps, the standard PRA requires incubation, followed by chemical fixation, crystal violet staining, and manual counting to yield a single endpoint readout. In contrast, the automated label-free PRA presented in this work continuously acquires time-lapse images during incubation (i.e., within an incubator). Following differential image processing, a neural network-based algorithm analyzes these temporal differential image stacks to generate detection masks and extract key quantitative metrics—e.g., total PFU count $N$, total PFU area coverage $A$, new PFU formation $\Delta N$, and incremental PFU area growth $\Delta A$—starting as early as 22 hours post-infection. **(c)** A comprehensive qualitative comparison highlighting the proposed platform's advantages over standard PRA in terms of time efficiency, labor reduction, safety, high-throughput capability, lower cost, and label-free dynamic monitoring and quantification of PFUs.

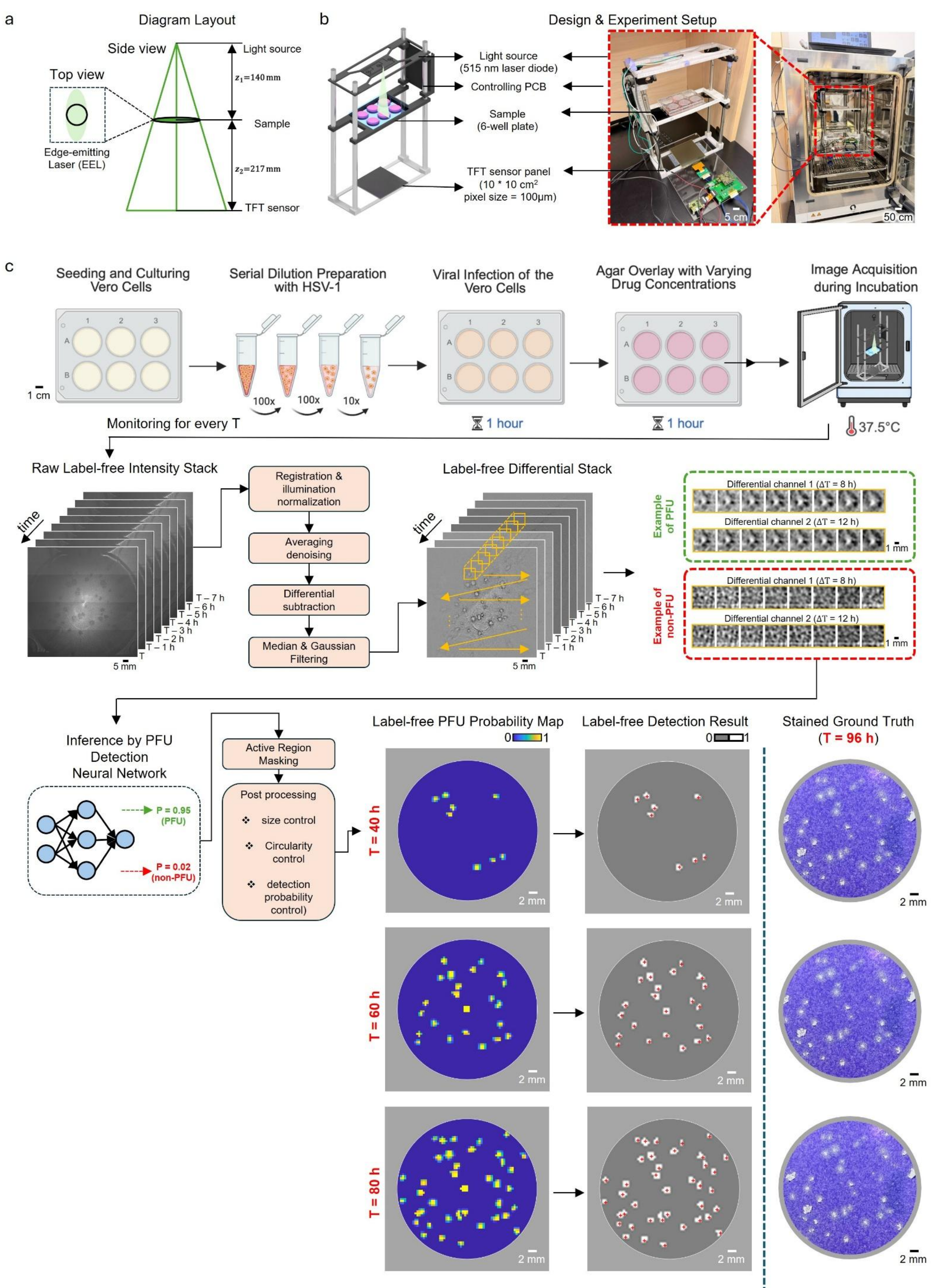

**Figure 2. Hardware design and computational processing pipeline of our label-free, rapid, and automated PRA. (a)** Schematic layout of our customized lens-free imaging setup, utilizing a 515 nm laser

diode as the light source and a TFT image sensor with an ultra-large active area (100 $cm^2$, pixel size: 100 μm), which corresponds to a field-of-view (FOV) of ~15.4 $cm^2$ at the sample plane. **(b)** Computer-aided design (CAD) model and photographs of our setup, shown both outside and operating within a standard incubator. **(c)** The workflow of our system. Following standard PRA sample preparation, raw lens-free images are captured via time-lapse imaging over 96 hours at 1-hour intervals. These raw intensity images undergo registration, illumination normalization, denoising, and differential subtraction to generate the label-free differential stacks with $\Delta T$ = 8 h and $\Delta T$ = 12 h. These differential stacks contain unique spatio-temporal information highlighting the distinct dynamic signatures of emerging PFUs from non-PFU artifacts and are then fed into a PFU detection neural network. Subsequent post-processing generates label-free PFU probability maps and binary detection results at continuous time points (examples at 40, 60, and 80 hours of incubation are shown here), which exhibit high concordance with the corresponding traditional stained ground truth at T = 96 h.

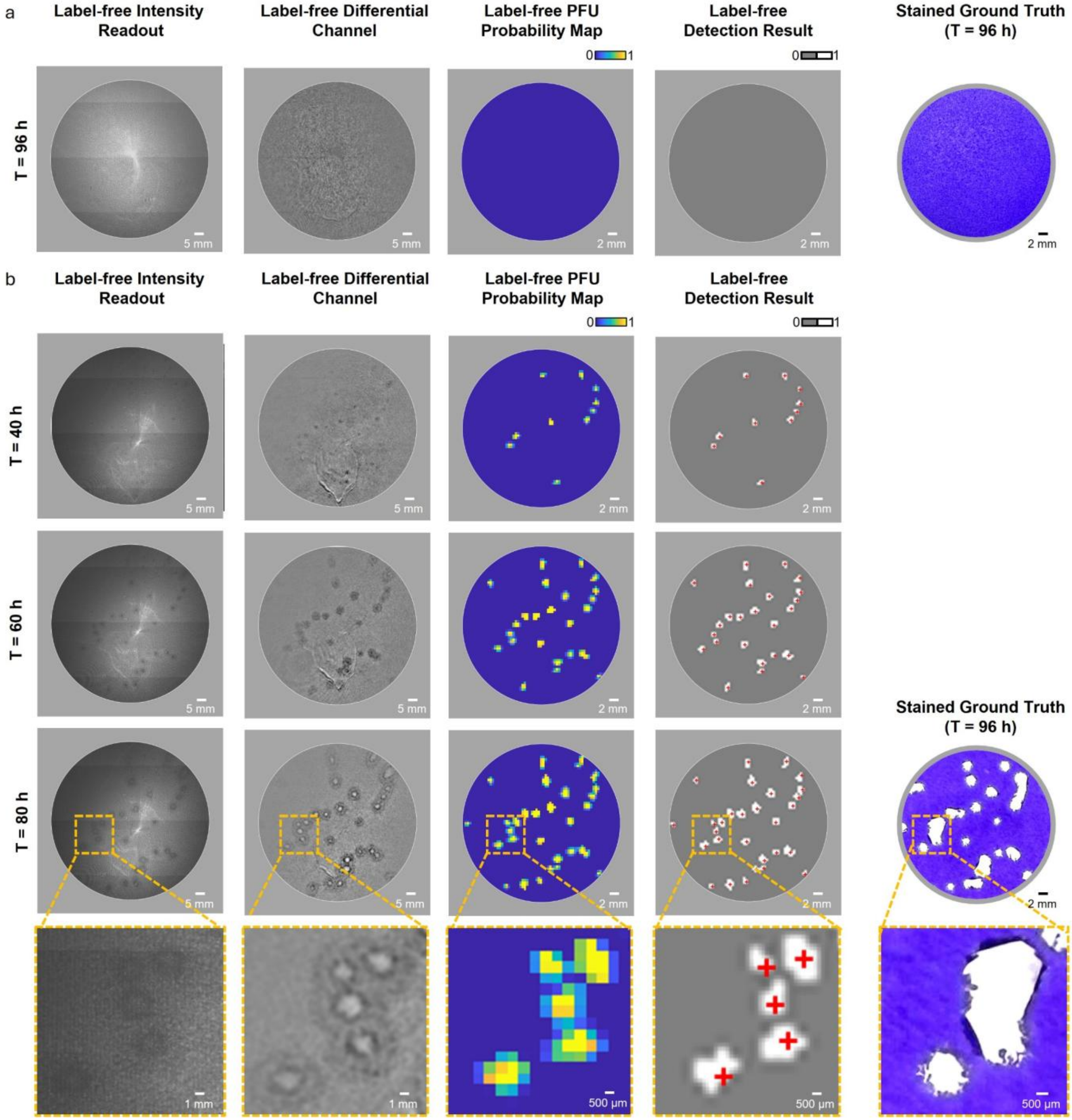

**Figure 3. PFU evaluation on untreated, label-free samples. (a)** Evaluation of a representative uninfected negative control well at 96 hours. The label-free intensity readout and differential channel ($\Delta T = 12$ h) are presented in the 1st and 2nd columns. The label-free probability map and detection result (the 3rd and 4th columns) yield zero false-positive detections, matching the chemically stained ground truth at 96 hours (the 5th column). Notably, this complete absence of false positives is consistently maintained across all time points and all negative control samples. **(b)** Time-lapse monitoring of a representative HSV-1-infected, ACV-untreated positive control well at 40, 60, and 80 h post-infection. The label-free probability maps and detection results (the 3rd and 4th columns) accurately predict PFU locations, exhibiting strong spatial concordance with the chemically stained ground truth at 96 hours. Crucially, the zoomed-in regions (yellow boxes) highlight the system's ability to resolve complex, overlapping PFU clusters. Owing to its early detection capabilities, 4 distinct PFUs are successfully identified at 80 h, whereas they coalesce into a single indistinguishable plaque in the traditional 96-h stained ground truth. Note that the different scale bars across channels arise from the system's 2.55× fringe magnification factor: raw lens-free intensity images and differential images are defined in the sensor plane, whereas the probability maps and detection masks are mapped back to the sample plane.

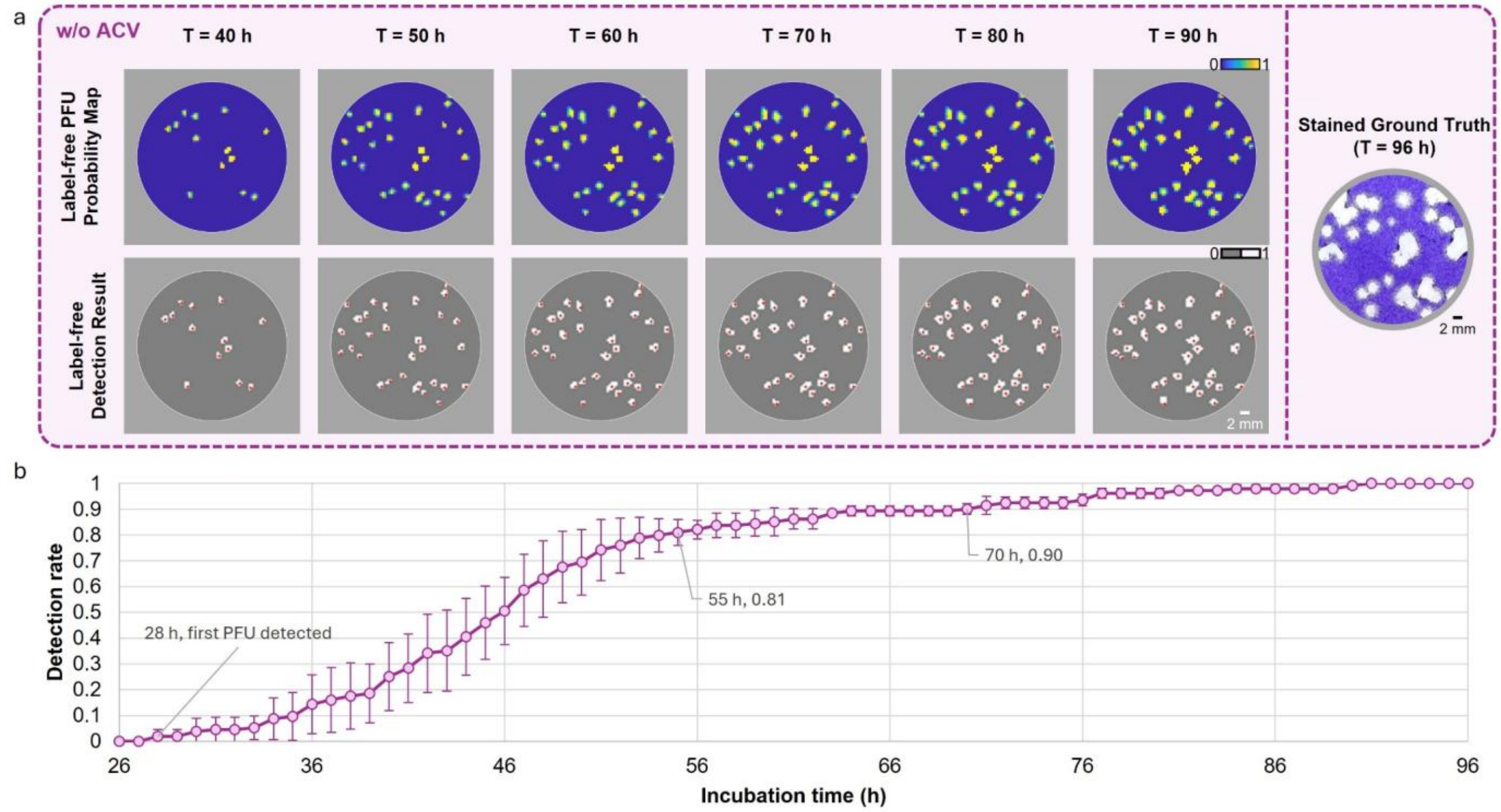

**Figure 4. Quantitative PFU analysis on ACV-untreated, label-free samples. (a)** Time-lapse monitoring of a representative HSV-1-infected, ACV-untreated well (positive control). The label-free PFU probability maps (top row) and final detection results (bottom row) are presented at 10-hour intervals from 40 h to 90 h. These computational outputs accurately map the expanding PFUs, culminating in high spatial agreement with the 96-hour chemically stained ground truth. **(b)** PFU detection rate vs. the incubation time. The detection rate was defined as the number of system-detected PFUs divided by the total PFU count confirmed via the 96-hour-stained ground truth. The system achieves its first successful PFU detection as early as 28 h post-infection. The detection rate steadily climbs alongside viral progression, reaching 81% at 55 hours and attaining 90% by 70 hours. Error bars represent the standard deviation across 3 replicates.

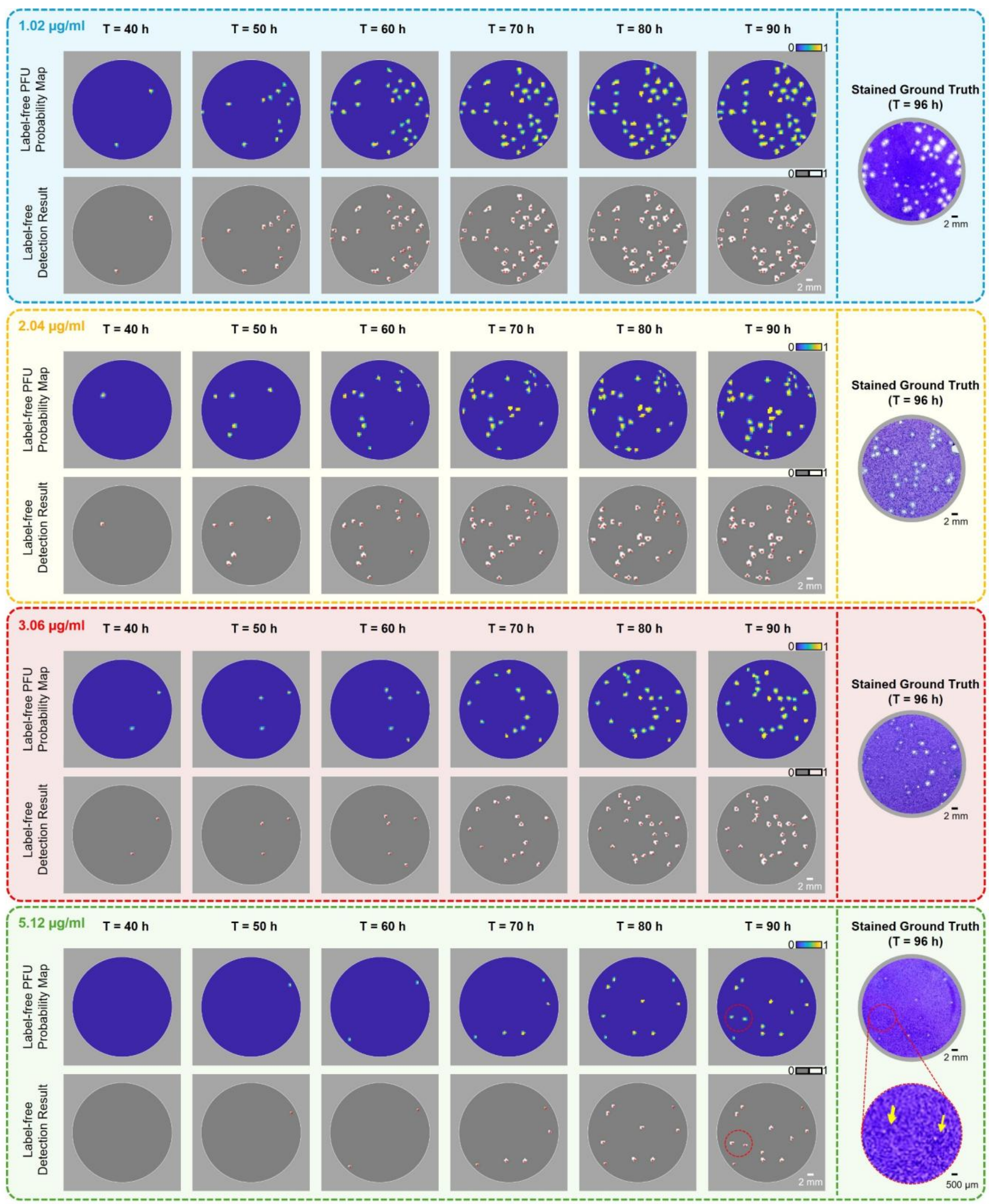

**Figure 5. PFU evaluation on ACV-treated, label-free samples.** Time-lapse monitoring (sampled at 10-hour intervals from 40 h to 90 h) of representative HSV-1-infected wells treated with varying concentrations of ACV (1.02, 2.04, 3.06, and 5.12 µg/mL). The label-free probability maps and detection results illustrate a clear dose-dependent antiviral effect, characterized by delayed initial PFU emergence, reduced total PFU

counts, and restricted spatial expansion. Notably, at the highest dose (5.12 μg/mL), the algorithm successfully identifies heavily suppressed, small-scale PFUs (indicated by red dashed circles).

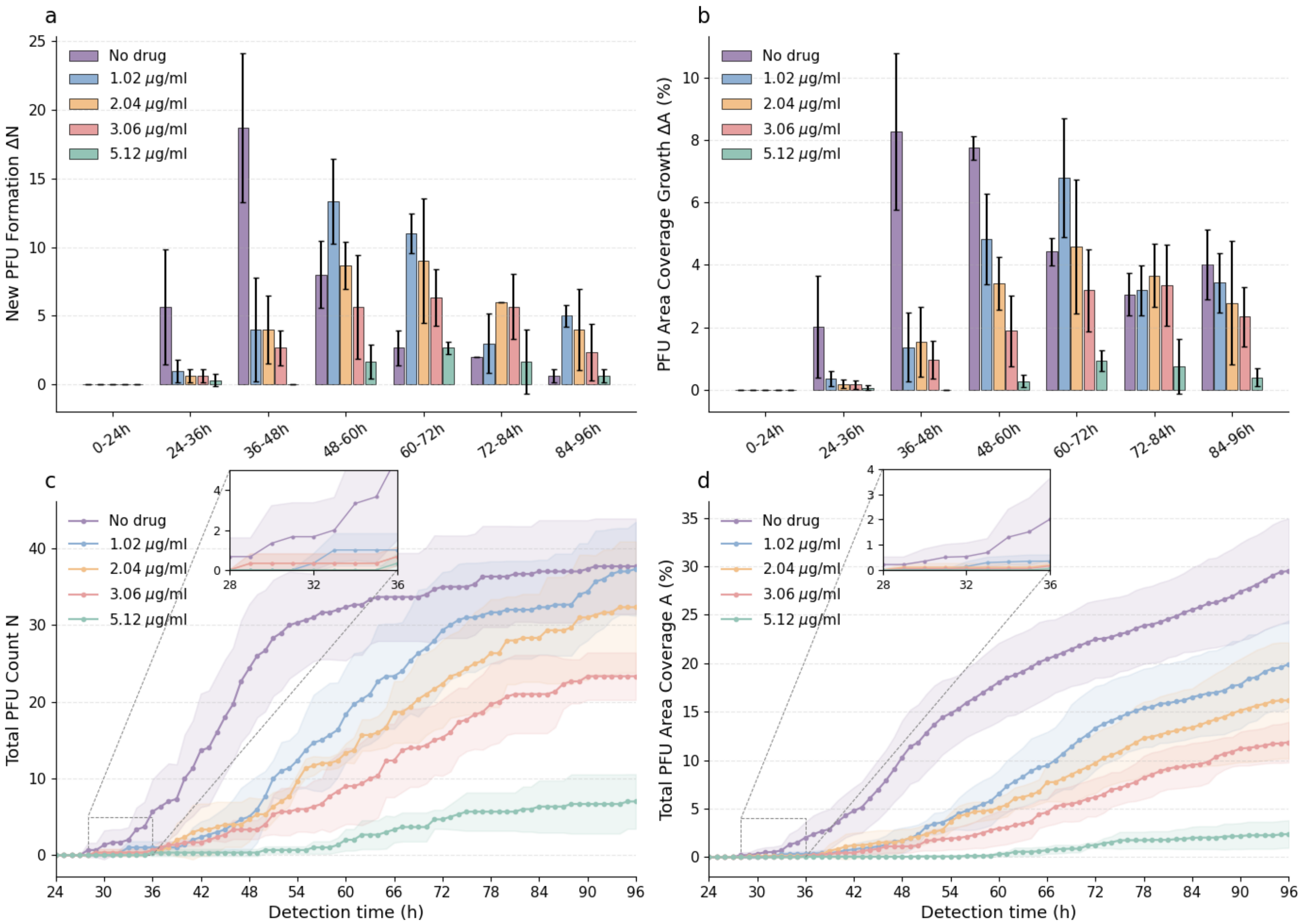

**Figure 6. Quantitative PFU analysis on ACV-treated, label-free samples. (a)** New PFU formation ($\Delta N$) was evaluated over consecutive 12-hour intervals (except for an initial 0–24 h incubation period) for untreated ("no drug") and ACV-treated samples (with drug concentrations of 1.02, 2.04, 3.06, and 5.12 μg/mL). **(b)** The corresponding PFU area coverage growth percentage ($\Delta A$) was calculated over the same time intervals. Error bars in (a) and (b) represent the standard deviation (n = 3). **(c)** The total PFU count ($N$) vs. detection time throughout the incubation. **(d)** The total PFU area coverage percentage ($A$). In (c) and (d), solid lines indicate the mean values, and shaded regions denote the standard deviation (n = 3). The zoomed-in plots highlight the system's early detection sensitivity between 28 and 36 hours. Together, the profiles reveal a distinct dose-dependent delay and suppression in both plaque emergence and cytopathic expansion.

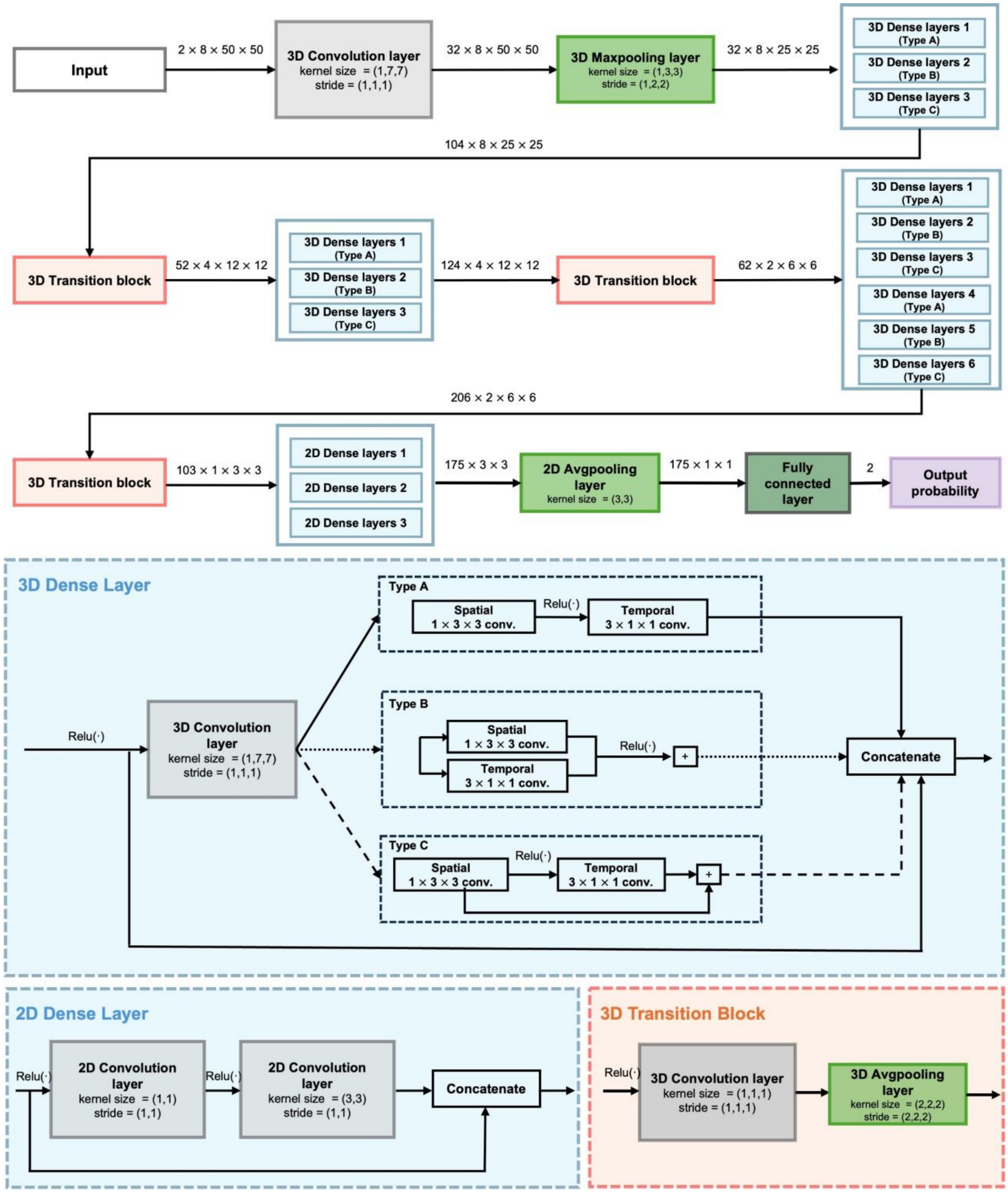

**Figure 7. Network architecture of the PFU detection neural network.**

## REFERENCES

1. Saindane, R. A. & Pathania, A. Targeting Key Stages of the Viral Entry and Life Cycle: A Comprehensive Overview of the Mechanisms of Antiviral Actions. *Methods Mol Biol* **2927**, 259-286 (2025). https://doi.org/10.1007/978-1-0716-4546-8_15
2. Kausar, S. *et al.* A review: Mechanism of action of antiviral drugs. *Int J Immunopathol Pharmacol* **35**, 20587384211002621 (2021). https://doi.org/10.1177/20587384211002621
3. Li, Xinlei & Peng, Tao. Strategy, Progress, and Challenges of Drug Repurposing for Efficient Antiviral Discovery. *Frontiers in Pharmacology* **Volume 12 - 2021** (2021). https://doi.org/10.3389/fphar.2021.660710
4. Adamson, Catherine S. *et al.* Antiviral drug discovery: preparing for the next pandemic. *Chemical Society Reviews* **50**, 3647-3655 (2021). https://doi.org/10.1039/D0CS01118E
5. Namba-Nzanguim, Cyril T. *et al.* Artificial intelligence for antiviral drug discovery in low resourced settings: A perspective. *Frontiers in Drug Discovery* **Volume 2 - 2022** (2022). https://doi.org/10.3389/fddsv.2022.1013285
6. Magden, J., Kääriäinen, L. & Ahola, T. Inhibitors of virus replication: recent developments and prospects. *Appl Microbiol Biotechnol* **66**, 612-621 (2005). https://doi.org/10.1007/s00253-004-1783-3
7. Rumlová, M. & Ruml, T. In vitro methods for testing antiviral drugs. *Biotechnol Adv* **36**, 557-576 (2018). https://doi.org/10.1016/j.biotechadv.2017.12.016
8. FDA. *Guidance for Industry Antiviral Product Development - Conducting and Submitting Virology Studies to the Agency*, <https://www.fda.gov/media/71223/download> (2006).
9. Landry, M. L. *et al.* A standardized plaque reduction assay for determination of drug susceptibilities of cytomegalovirus clinical isolates. *Antimicrob Agents Chemother* **44**, 688-692 (2000). https://doi.org/10.1128/aac.44.3.688-692.2000
10. Baer, A. & Kehn-Hall, K. Viral concentration determination through plaque assays: using traditional and novel overlay systems. *J Vis Exp*, e52065 (2014). https://doi.org/10.3791/52065
11. Yin, Y. *et al.* Screening and verification of antiviral compounds against HSV-1 using a method based on a plaque inhibition assay. *BMC Infect Dis* **23**, 890 (2023). https://doi.org/10.1186/s12879-023-08843-3
12. Hu, Y., Ma, C. & Wang, J. Cytopathic Effect Assay and Plaque Assay to Evaluate in vitro Activity of Antiviral Compounds Against Human Coronaviruses 229E, OC43, and NL63. *Bio Protoc* **12**, e4314 (2022). https://doi.org/10.21769/BioProtoc.4314
13. Yin, Yingxian *et al.* Screening and verification of antiviral compounds against HSV-1 using a method based on a plaque inhibition assay. *BMC Infectious Diseases* **23**, 890 (2023). https://doi.org/10.1186/s12879-023-08843-3
14. Wang, H. Practical updates in clinical antiviral resistance testing. *J Clin Microbiol* **62**, e0072823 (2024). https://doi.org/10.1128/jcm.00728-23
15. Liu, Tairan *et al.* Rapid and stain-free quantification of viral plaque via lens-free holography and deep learning. *Nature Biomedical Engineering* **7**, 1040-1052 (2023). https://doi.org/10.1038/s41551-023-01057-7
16. Thi, Thuong Nguyen *et al.* Rapid determination of antiviral drug susceptibility of herpes simplex virus types 1 and 2 by real-time PCR. *Antiviral Research* **69**, 152-157 (2006). https://doi.org/https://doi.org/10.1016/j.antiviral.2005.11.004

17 Piret, J., Goyette, N. & Boivin, G. Novel Method Based on Real-Time Cell Analysis for Drug Susceptibility Testing of Herpes Simplex Virus and Human Cytomegalovirus. *J Clin Microbiol* **54**, 2120-2127 (2016). https://doi.org/10.1128/jcm.03274-15

18 Stránská, R., van Loon, A. M., Polman, M. & Schuurman, R. Application of real-time PCR for determination of antiviral drug susceptibility of herpes simplex virus. *Antimicrob Agents Chemother* **46**, 2943-2947 (2002). https://doi.org/10.1128/aac.46.9.2943-2947.2002

19 Virók, D. P. *et al.* A direct quantitative PCR-based measurement of herpes simplex virus susceptibility to antiviral drugs and neutralizing antibodies. *J Virol Methods* **242**, 46-52 (2017). https://doi.org/10.1016/j.jviromet.2017.01.007

20 Standring-Cox, R., Bacon, T. H. & Howard, B. A. Comparison of a DNA probe assay with the plaque reduction assay for measuring the sensitivity of herpes simplex virus and varicella-zoster virus to penciclovir and acyclovir. *J Virol Methods* **56**, 3-11 (1996). https://doi.org/10.1016/0166-0934(95)01889-1

21 Safrin, S., Phan, L. & Elbeik, T. A comparative evaluation of three methods of antiviral susceptibility testing of clinical herpes simplex virus isolates. *Clin Diagn Virol* **4**, 81-91 (1995). https://doi.org/10.1016/0928-0197(94)00059-4

22 Danve, C., Morfin, F., Thouvenot, D. & Aymard, M. A screening dye-uptake assay to evaluate in vitro susceptibility of herpes simplex virus isolates to acyclovir. *J Virol Methods* **105**, 207-217 (2002). https://doi.org/10.1016/s0166-0934(02)00103-9

23 Payne, A. F., Binduga-Gajewska, I., Kauffman, E. B. & Kramer, L. D. Quantitation of flaviviruses by fluorescent focus assay. *J Virol Methods* **134**, 183-189 (2006). https://doi.org/10.1016/j.jviromet.2006.01.003

24 Masci, A. L. *et al.* Integration of Fluorescence Detection and Image-Based Automated Counting Increases Speed, Sensitivity, and Robustness of Plaque Assays. *Mol Ther Methods Clin Dev* **14**, 270-274 (2019). https://doi.org/10.1016/j.omtm.2019.07.007

25 Fabiani, M., Limongi, D., Palamara, A. T., De Chiara, G. & Marcocci, M. E. A Novel Method to Titrate Herpes Simplex Virus-1 (HSV-1) Using Laser-Based Scanning of Near-Infrared Fluorophores Conjugated Antibodies. *Front Microbiol* **8**, 1085 (2017). https://doi.org/10.3389/fmicb.2017.01085

26 Praditya, D. F., Waluyo, D. & Nozaki, T. Reporter-expressing viruses for antiviral drug discovery research. *Front Cell Infect Microbiol* **15**, 1645104 (2025). https://doi.org/10.3389/fcimb.2025.1645104

27 Zhu, Qin-Chang, Wang, Yi & Peng, Tao. Herpes Simplex Virus (HSV) Immediate-Early (IE) Promoter-Directed Reporter System for the Screening of Antiherpetics Targeting the Early Stage of HSV Infection. *SLAS Discovery* **15**, 1016-1020 (2010). https://doi.org/https://doi.org/10.1177/1087057110372804

28 Srimathi, Siddharth Raghu *et al.* Microfluidic digital focus assays for the quantification of infectious influenza virus. *Lab on a Chip* **25**, 2004-2016 (2025). https://doi.org/10.1039/D4LC00940A

29 Wei, Qingshan *et al.* Fluorescent Imaging of Single Nanoparticles and Viruses on a Smart Phone. *ACS Nano* **7**, 9147-9155 (2013). https://doi.org/10.1021/nn4037706

30 Arias-Arias, J. L., Corrales-Aguilar, E. & Mora-Rodríguez, R. A. A Fluorescent Real-Time Plaque Assay Enables Single-Cell Analysis of Virus-Induced Cytopathic Effect by Live-Cell Imaging. *Viruses* **13** (2021). https://doi.org/10.3390/v13071193

31 van Remmerden, Yvonne *et al.* An improved respiratory syncytial virus neutralization assay based on the detection of green fluorescent protein expression and automated plaque counting. *Virology Journal* **9**, 253 (2012). https://doi.org/10.1186/1743-422X-9-253

32 Kühn, J. *et al.* Label-free cytotoxicity screening assay by digital holographic microscopy. *Assay Drug Dev Technol* **11**, 101-107 (2013). https://doi.org/10.1089/adt.2012.476

33 Lebourgeois, Samuel *et al.* Development of a Real-Time Cell Analysis (RTCA) Method as a Fast and Accurate Method for Detecting Infectious Particles of the Adapted Strain of Hepatitis A Virus. *Frontiers in Cellular and Infection Microbiology* **Volume 8 - 2018** (2018). https://doi.org/10.3389/fcimb.2018.00335

34 Amarilla, A. A. *et al.* An Optimized High-Throughput Immuno-Plaque Assay for SARS-CoV-2. *Front Microbiol* **12**, 625136 (2021). https://doi.org/10.3389/fmicb.2021.625136

35 Wen, Z. *et al.* Development and application of a higher throughput RSV plaque assay by immunofluorescent imaging. *J Virol Methods* **263**, 88-95 (2019). https://doi.org/10.1016/j.jviromet.2018.10.022

36 Han, C. & Yang, C. Viral plaque analysis on a wide field-of-view, time-lapse, on-chip imaging platform. *Analyst* **139**, 3727-3734 (2014). https://doi.org/10.1039/c3an02323k

37 Vanhulle, Emiel *et al.* Cellular electrical impedance to profile SARS-CoV-2 fusion inhibitors and to assess the fusogenic potential of spike mutants. *Antiviral Research* **213**, 105587 (2023). https://doi.org/https://doi.org/10.1016/j.antiviral.2023.105587

38 Oeyen, M., Meyen, E., Doijen, J. & Schols, D. In-Depth Characterization of Zika Virus Inhibitors Using Cell-Based Electrical Impedance. *Microbiol Spectr* **10**, e0049122 (2022). https://doi.org/10.1128/spectrum.00491-22

39 de Oliveira, A., Prince, D., Lo, C. Y., Lee, L. H. & Chu, T. C. Antiviral activity of theaflavin digallate against herpes simplex virus type 1. *Antiviral Res* **118**, 56-67 (2015). https://doi.org/10.1016/j.antiviral.2015.03.009

40 Beyleveld, Grant, White, Kris M., Ayllon, Juan & Shaw, Megan L. New-generation screening assays for the detection of anti-influenza compounds targeting viral and host functions. *Antiviral Research* **100**, 120-132 (2013). https://doi.org/https://doi.org/10.1016/j.antiviral.2013.07.018

41 Petkidis, A., Andriasyan, V., Murer, L., Volle, R. & Greber, U. F. A versatile automated pipeline for quantifying virus infectivity by label-free light microscopy and artificial intelligence. *Nat Commun* **15**, 5112 (2024). https://doi.org/10.1038/s41467-024-49444-1

42 Cacciabue, M., Currá, A. & Gismondi, M. I. ViralPlaque: a Fiji macro for automated assessment of viral plaque statistics. *PeerJ* **7**, e7729 (2019). https://doi.org/10.7717/peerj.7729

43 Yakimovich, A. *et al.* Plaque2.0-A High-Throughput Analysis Framework to Score Virus-Cell Transmission and Clonal Cell Expansion. *PLoS One* **10**, e0138760 (2015). https://doi.org/10.1371/journal.pone.0138760

44 Ali, M. *et al.* How Deep Learning in Antiviral Molecular Profiling Identified Anti-SARS-CoV-2 Inhibitors. *Biomedicines* **11** (2023). https://doi.org/10.3390/biomedicines11123134

45 Mudanyali, Onur *et al.* Compact, light-weight and cost-effective microscope based on lensless incoherent holography for telemedicine applications. *Lab on a Chip* **10**, 1417-1428 (2010). https://doi.org/10.1039/C000453G

46 Greenbaum, Alon *et al.* Imaging without lenses: achievements and remaining challenges of wide-field on-chip microscopy. *Nature Methods* **9**, 889-895 (2012). https://doi.org/10.1038/nmeth.2114

47 Ozcan, Aydogan & McLeod, Euan. Lensless Imaging and Sensing. *Annual Review of Biomedical Engineering* **18**, 77-102 (2016). https://doi.org/https://doi.org/10.1146/annurev-bioeng-092515-010849

48 Huang, Gao, Liu, Zhuang & Weinberger, Kilian Q. Densely Connected Convolutional Networks. *2017 IEEE Conference on Computer Vision and Pattern Recognition (CVPR)*, 2261-2269 (2016).

49 Qiu, Z., Yao, T. & Mei, T. in *2017 IEEE International Conference on Computer Vision (ICCV).* 5534-5542.

50 Kakizoe, Y. *et al.* A method to determine the duration of the eclipse phase for in vitro infection with a highly pathogenic SHIV strain. *Sci Rep* **5**, 10371 (2015). https://doi.org/10.1038/srep10371

51 Rivenson, Yair *et al.* Virtual histological staining of unlabelled tissue-autofluorescence images via deep learning. *Nature Biomedical Engineering* **3**, 466-477 (2019). https://doi.org/10.1038/s41551-019-0362-y

52 Bai, Bijie *et al.* Label-Free Virtual HER2 Immunohistochemical Staining of Breast Tissue using Deep Learning. *BME Frontiers* **2022**, 9786242 (2022). https://doi.org/doi:10.34133/2022/9786242